\documentclass{aa}  

\usepackage{graphicx}
\usepackage{txfonts}
\usepackage[normalem]{ulem}

\begin{document}

   \title{Estimating the longitudinal magnetic field in the chromosphere of quiet-Sun magnetic concentrations}

   \author{S. Esteban Pozuelo\inst{1,2} \and  A. Asensio Ramos\inst{1,2} \and J. de la Cruz Rodríguez\inst{3} \and J. Trujillo Bueno\inst{1,2,4} \and M.~J. Martínez González\inst{1,2}}

   \institute{Instituto de Astrof\'{i}sica de Canarias, E-38205 La Laguna, Tenerife, Spain \\
              \email{sesteban@iac.es}
         \and
             Universidad de La Laguna, Dept. Astrof\'{i}sica, E-38206 La Laguna, Tenerife, Spain
                \and
                        Institute for Solar Physics, Dept. of Astronomy, Stockholm University, AlbaNova University Center, SE-106 91 Stockholm, Sweden
                \and
                        Consejo Superior de Investigaciones Cient\'{i}ficas, Spain
                        }

   \date{Received ; accepted}
 
  \abstract
   {Details of the magnetic field in the quiet Sun chromosphere are key to our understanding of essential aspects of the solar atmosphere. However, the strength and orientation of this magnetic field have not been thoroughly studied at high spatial resolution.}
   {We aim to determine the longitudinal magnetic field component ($B_{\|}$) of quiet Sun regions depending on their size.}
   {We estimated $B_{\|}$ by applying the weak-field approximation (WFA) to high-spatial-resolution \ion{Ca}{II} 854.2~nm data taken with the Swedish 1m Solar Telescope. Specifically, we analyzed the estimates inferred for different spectral ranges using the data at the original cadence and temporally integrated signals.}
   {The longitudinal magnetic field in each considered plasma structure correlates with its size. Using a spectral range restricted to the line core leads to chromospheric longitudinal fields varying from $\sim$50~G at the edges to 150--500~G at the center of the structure. These values increase as the spectral range widens due to the photospheric contribution. However, the difference between this contribution and the chromospheric one is not uniform for all structures. Small and medium-sized concentrations show a steeper height gradient in $B_{\|}$ compared to their chromospheric values, so estimates for wider ranges are less trustworthy. Signal addition does not alleviate this situation as the height gradients in $B_{\|}$ are consistent with time. Finally, despite the amplified noise levels that deconvolving processes may cause, data restored with the destretching technique show similar results, though are affected by smearing.}
   {We obtained $B_{\|}$ estimates similar to those previously found, except for large concentrations and wide spectral ranges. In addition, we report a correlation between the height variation of $B_{\|}$ compared to the chromospheric estimates and the concentration size. This correlation affects the difference between the photospheric and chromospheric magnetic flux values and the reliability of the estimates for wider spectral ranges.}


   \maketitle
%

\section{Introduction}
\label{sec:intro}

Magnetic fields play a crucial role in explaining diverse processes throughout the solar atmosphere. However, the physical conditions in the chromosphere hinder the study of magnetic fields in this layer, which is key for linking the photospheric and coronal field topologies.

Polarization measurements in the photosphere of the quiet Sun show magnetic elements of diverse sizes and lifetimes. These structures are divided into a network \citep{1967SoPh....1..171S} and an internetwork (\citealt{1975BAAS....7..346L}; \citealt{1975BAAS....7R.346S}). The former consists of nearly vertical magnetic flux tubes expanding with height due to force imbalances. Network structures are bright due to their lower opacity \citep{1976SoPh...50..269S} and harbor kG magnetic fields. In contrast, the internetwork is formed by small and short-lived structures randomly wandering due to convective motions until they merge with the network (e.g., \citealt{1985AuJPh..38..961Z}; \citealt{2007ApJ...666L.137C}; \citealt{2009ApJ...700.1391M}; \citealt{2014ApJ...797...49G}). These structures often appear as bipolar features linked by magnetic loops (e.g., \citealt{2007ApJ...666L.137C}; \citealt{2009ApJ...700.1391M}). Internetwork fields have strengths of the order of hG (e.g., \citealt{2003A&A...408.1115K}; \citealt{2007ApJ...659..829A}; \citealt{2007A&A...469L..39M}; \citealt{2016A&A...593A..93D}) and are crucial to our understanding of the quiet Sun magnetism.

The coupling of the quiet Sun magnetic field between the photosphere and the outer atmosphere is often described by umbrella-shaped canopies rising from network structures and expanding higher up. However, the inference of hG fields in the internetwork led to more complex schemes, such as the multiscale carpet model \citep{2003ApJ...597L.165S} where low-lying canopies link network and internetwork. In addition, the omnipresence of the internetwork suggests that it may contain an important fraction of the photospheric magnetic flux, thereby contributing greatly to the energy balance of the solar atmosphere \citep{2004Natur.430..326T}. In this regard, \citet{2014ApJ...797...49G} found internetwork magnetic flux values resembling those in the network and active regions and highlighted the importance of the internetwork as a contributor to the network flux. Moreover, internetwork fields are able to travel higher up across the photosphere \citep[e.g., ][]{2010ApJ...714L..94M, 2013A&A...556A...7G}. 

Despite great efforts made over the years, the magnetism of the quiet Sun photosphere is not totally understood due to the small scale and weakness of the signals of the observed magnetic elements (see the reviews by \citealp{2006ASPC..358..269T}, \citealt{2011ASPC..437..451S}, and \citealt{2019LRSP...16....1B}). These drawbacks are exacerbated in the chromosphere, where the particle density drops. This decrease benefits the expansion of magnetic fields leading to weak polarization signals. For instance, Stokes~$V$ amplitudes in the \ion{Ca}{II} 854.2~nm line in the quiet Sun are of about 10$\mathrm{^{-2}}$--10$\mathrm{^{-3}}$ relative to the continuum intensity. However, noise levels below 10$\mathrm{^{-3.5}}$ are required to detect linear polarization induced by the Zeeman effect in the $\lambda$8542 line \citep{2012A&A...543A..34D}. For field strengths of between 10 and 100~G, the Hanle effect also contributes to the linear polarization of this line. In this last case, the expected Stokes $Q$ and $U$ signals at disk center are of the order of 10$^{-4}$ \citep{2010ApJ...722.1416M}. In addition, a low collisional rate causes a weak coupling of the radiation to the local physical conditions, which invalidates the assumption of local thermodynamic equilibrium (LTE).

Fortunately, the use of circular polarization measurements in \ion{Ca}{II} 854.2~nm has shed light on the magnetism of the quiet-Sun chromosphere. For example, the analysis of these measurements, combined with information in other spectral lines has allowed the study of the interaction of rising small-scale internetwork loops with network loops and the role of cancellations of opposite-polarity internetwork photospheric fields in the local chromospheric heating \citep{2018ApJ...857...48G,2021ApJ...911...41G}.

In fact, the chromospheric heating is another key aspect related to the magnetic field in the quiet-Sun chromosphere. Diverse mechanisms can heat the chromosphere \citep[e.g.,][]{1989ApJ...346L..37K, 2012ApJ...747...87K, 2018ApJ...862L..24P}, but it is the Ohmic current dissipation that greatly contributes in the low-mid chromosphere of bright points and plages \citep{2022arXiv220301688M}. This predominance in the chromospheric heating makes the inference of the chromospheric magnetic field in quiet-Sun magnetic elements an appealing task as part of further exploration.

Despite the availability of diverse nonLTE inversion codes, the weak-field approximation (WFA) is commonly used to estimate the chromospheric magnetic field because it requires little time and is relatively straightforward (see Sect.~\ref{subsec:wfa}). Specifically, this technique works well for measurements in the $\lambda$8542 line \citep{hammar2014, 2018ApJ...866...89C}. According to \citet{2018ApJ...866...89C}, the longitudinal field estimates are reliable for field strengths up to 1.2~kG, but are underestimated for noise levels of 10$\mathrm{^{-3}}$ in units of the continuum intensity.

Given its importance, we aim to characterize the magnetic field of quiet-Sun structures depending on their size using high-spatial-resolution data taken in the \ion{Ca}{II} 854.2~nm line with the Swedish 1m Solar Telescope \citep[SST;][]{2003SPIE.4853..341S}. In this study, we determined the longitudinal magnetic field by applying the WFA to different spectral ranges across the line. In addition, we analyzed the estimates obtained from data at the original cadence and temporally integrated signals that mimic longer scanning times.

\begin{figure*} 
\centerline{\includegraphics[trim = {0cm 0cm 0cm 0cm}, clip, height = 0.93\textheight]{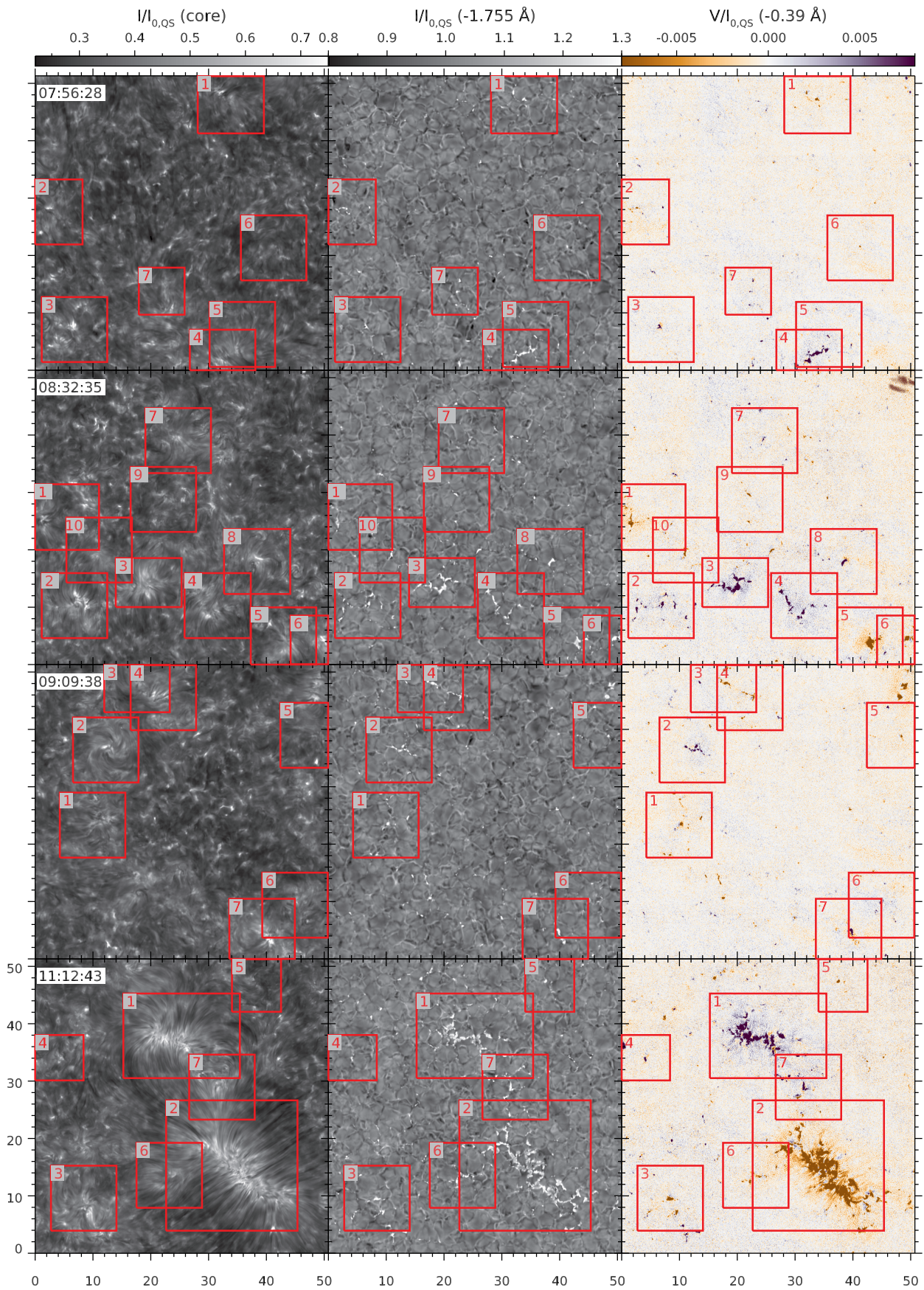}}
\caption{Intensity filtergrams of the \ion{Ca}{II} 854.2~nm line core and at $-$1.755~\AA \, (left and middle columns), and blue-wing magnetograms of \ion{Ca}{II} 854.2~nm at $-$0.39~\AA \, from the line center (right column). The red rectangles show the positions of the analyzed structures. The time stamp of each scan in UT is shown in the upper-left corner labels. Axes are labeled in arcseconds.}
\label{fig:data1}
\end{figure*}

\section{Observation and data reduction} \label{sec:observation}

On August 1, 2019, we acquired four time series of quiet-Sun regions at different positions over the center of the solar disk with the CRisp Imaging SpectroPolarimeter \citep[CRISP,][]{2006A&A...447.1111S, 2008ApJ...689L..69S} attached to the SST. Table~\ref{tab:description_regions} shows some observing characteristics of each sequence. Seeing conditions were good during the observing run except in the late morning. We therefore only analyzed the first 35 scans of the last time series.

Each time series consists of full-Stokes measurements in the \ion{Ca}{II} 854.2~nm line at high spatial resolution. The line was sampled at 21 nonequidistant wavelength positions between $\pm$1.040~\AA \, from the line center using steps of 65~m\AA \, in the core and wider in the wings, plus two extra points at $\pm$1.755~\AA. We acquired 12 accumulations per wavelength position. As the exposure time of each accumulation is 17~ms (with four modulation states), each sampling was completed every 31.5~s. The field of view (FOV) is $\sim$55\arcsec$\times$55\arcsec and the pixel size 0.057\arcsec.

\begin{table}
 \caption{Heliocentric coordinates of the FOV center at the beginning of the observing run, time interval, and number of scans of each dataset.}
\label{tab:description_regions}
\begin{tabular}{c c c c}     
 \hline \hline
Series & (x, y) & Start--End UT Time & Scans \\ 
\hline
1 & (1\arcsec, 0\arcsec) & 07:39--08:14 & 66 \\ 
2 & (2\arcsec, $-$58\arcsec) & 08:15--08:50 & 66 \\ 
3 & (9\arcsec, $-$122\arcsec) & 08:52--09:27 & 66 \\  
4 & (110\arcsec, 10\arcsec) & 10:55--11:25 & 57 \\ 
\hline
\end{tabular}
\end{table}

We performed the data reduction using the SSTRED pipeline (\citealt{2015A&A...573A..40D}; \citealp{2021A&A...653A..68L}). Images were restored with the Multi-Object-Multi-Frame-Blind-Deconvolution technique \citep[MOMFBD,][]{1994A&AS..107..243L, 2005SoPh..228..191V}. Finally, we performed the polarimetric calibration on a pixel-by-pixel basis as proposed by \citet{2008A&A...489..429V}. The noise level in the resulting Stokes~$Q$, $U$, and $V$ images is 2.3--2.8$\times$10$\mathrm{^{-3}}$, 2.4--2.6$\times$10$\mathrm{^{-3}}$, and 2.2--2.5$\times$10$\mathrm{^{-3}}$ as measured on the $Q$/$I_{0,QS}$, $U$/$I_{0,QS}$, and $V$/$I_{0,QS}$ maps at the first observed wavelength of each series. $I_{0,QS}$ refers to the first observed wavelength intensity averaged over a very quiet area. Given the weak linear polarization signals expected in the quiet-Sun chromosphere (see Sect.~\ref{sec:intro}), the polarimetric sensitivity of our spectropolarimetric data is not large enough to detect them in all the studied elements.

\section{Analysis} \label{sec:analysis}

Considering their brightness and conspicuous Stokes~$V$ signals, we search for field concentrations in the \ion{Ca}{II} 854.2~nm intensity filtergrams at $-$1.755~\AA\, and magnetograms at $\pm$0.39~\AA \, with the CRISPEX tool \citep{2012ApJ...750...22V}. We found 31 concentrations of different sizes, which we delimited by adopting thresholds of intensity at --1.755~\AA \, and circular polarization at --0.39~\AA\, of 0.95~$I_{0,QS}$ and $\pm$4$\times$10$\mathrm{^{-3}}$~$I_{0,QS}$. The red rectangles overplotted in Fig.~\ref{fig:data1} display the location of these structures. Most of the concentrations are visible during the entire run and often undergo fragmentations and mergings with other nearby concentrations. Such interactions are also included in this study.

While static plane-parallel mediums show antisymmetric Stokes~$V$ profiles, asymmetries appear in realistic model atmospheres. These asymmetries are mainly due to velocity gradients along the line of sight (LOS) and can be enhanced by magnetic-field gradients with height (\citealt{1975A&A....41..183I}; \citealt{1978A&A....64...67A}). Additionally, the presence of different \ion{Ca}{II} isotopes in the solar atmosphere also induces asymmetries in the $\lambda$8542 line \citep{2014ApJ...784L..17L}. We used an automated code to identify pixels with two-lobed $V$ profiles. Before that, we corrected the circular polarization signals for residual crosstalk from Stokes~$I$ similarly to \citet{2014ApJ...781..126O}, resulting in noise levels of 2.1--2.4$\times$10$\mathrm{^{-3}}$ as measured on the $V$/$I_{0,QS}$ maps at the first observed wavelength of each sequence. We also applied a principal component analysis denoising technique (PCA, \citealt{1901Pearson}, \citealt{1933Hotelling}) to the original circular polarization signals, leading to better-defined $V$ profiles that prevent detection errors. Specifically, we used four eigenvectors for the Stokes $V$ profiles in the first and third series and five in the second and fourth ones. We also smoothed the $V$ profiles with a boxcar average of a width of 2 pixels to ease the detection of the desired pixels. Finally, the subsequent analysis was performed using the signals resulting from the application of the PCA technique.

In addition to using the data at the original cadence, we analyzed the results from different scanning times ($\mathrm{\Delta t}$) to inspect the variation of the longitudinal magnetic field (see Sect.~\ref{subsec:wfa}) with the signal addition, in the hope of increasing the detectability of weaker fields. For this aim, we first identified the scans showing each structure and grouped them as sets of two, four, and ten scans. Then, we combined the two-lobed Stokes~$V$ signals and the corresponding Stokes~$I$ profiles at each pixel according to each set of scans to imitate observations with $\mathrm{\Delta t}$ of 1.05, 2.10, and 5.25 minutes. We also performed these combinations using all scans displaying each structure. We ruled out pixels with complex $V$~profiles by considering that their signals are zero. To avoid signal cancellation, we also discarded pixels with reversed polarities compared to that prevailing in each structure.

\subsection{Application of the WFA}
\label{subsec:wfa}

The WFA can be applied when the Zeeman splitting is much smaller than the Doppler width of the spectral line under consideration and the longitudinal component of the magnetic field is constant along the LOS \citep[see Sect. 9.6 in][]{2004ASSL..307.....L}. Under these circumstances, the observed Stokes~$V$ profile is related to the observed Stokes~$I$ profile as

\begin{equation}
V(\lambda) \ = \ -C \ f \ B \ \cos\gamma \ \frac{\partial I(\lambda)}{\partial\lambda},
\label{eq:wfa_1}
\end{equation}

\noindent with $C\,=\,4.6686\times10^{-13}\,\lambda_{0}^{2}\,g_{\mathrm{eff}}$, where $\lambda_{0}$ and $g_\mathrm{eff}$ are the central wavelength (in \AA) and the effective Land\'{e} factor of the line, respectively. The angle $\gamma$ is the inclination of the magnetic field with respect to the LOS direction. The longitudinal component of the magnetic field vector is $B_{\|}$ = $B \,\cos\gamma$, with $B$ the field strength (in G). The observed profile $I(\lambda)$ is assumed to be the intensity that would emerge with zero magnetic field. The filling factor $f$ is the fractional area occupied by the magnetic component within the resolution element. For simplicity, we assume $f$~=~1. Thus, the shape of Stokes~$V$ is proportional to the derivative of Stokes~$I$ scaled by the $B_{\|}$ estimate.

We compute $B_{\|}$ as the maximum-likelihood estimation from Eq.~(\ref{eq:wfa_1}) assuming
Gaussian noise with zero mean and variance $\sigma^2$, which leads to

\begin{equation}
B_{\|} \ = \ - \ \dfrac{\sum\limits_{i} \ \frac{\partial I(\lambda_{i})}{\partial\lambda_{i}} \ V(\lambda_{i})}{C \ \sum\limits_{i}\left(\frac{\partial I(\lambda_{i})}{\partial\lambda_{i}}\right)^{2}}.
\label{eq:wfa_2}
\end{equation}

\subsection{Definition of the analyzed spectral ranges}

The \ion{Ca}{II} 854.2~nm line covers a large range of formation heights \citep[e.g.,][]{2006ASPC..354..313U, 2008A&A...480..515C, 2016MNRAS.459.3363Q}. Its core is formed in the low- to mid-chromosphere, while its wings have a photospheric contribution. The variation from the LTE wings to the nonLTE chromospheric core occurs at $\pm$0.03--0.04 nm from the line center \citep{2008A&A...480..515C}. Therefore, the estimation of $B_{\|}$ values from the low- to mid-chromosphere is possible if the spectral range is restricted to the line core \citep{2015ApJ...801...16C}. Given its wide formation region, delimiting spectral ranges in the $\lambda$8542 line is a common method to infer $B_{\|}$ at certain heights \citep[e.g.,][]{2013A&A...556A.115D, 2019ApJ...870...88E, 2020A&A...642A.210M}. Recently, this approach was also applied to synthetic data in the \ion{Mg}{II}~h and k lines \citep{2022arXiv221114044A}. In this paper, we applied Eq.~(\ref{eq:wfa_2}) to different spectral ranges across the line to inspect how the $B_{\|}$ estimates differ. Specifically, we selected almost all the spectral sampling ($\Delta \lambda = \pm1.040$~\AA), a limited range including both lobes of the $V$~profile ($\pm0.325$~\AA), and the inner-core positions ($\pm0.130$~\AA). Hereafter, we denote these ranges as $\mathrm{\Delta \lambda_{1}}$, $\mathrm{\Delta \lambda_{2}}$, and $\mathrm{\Delta \lambda_{3}}$, respectively.

\section{Results} \label{sec:results}

The analyzed quiet-Sun structures evolve differently with time. While some of them undergo fragmentations, others either grow as they merge with nearby chunks or expand before decreasing. All cases show two-lobed $V$~profiles coinciding with Stokes~$I$ signals that sometimes show asymmetries.

\begin{figure} 
\centerline{\includegraphics[trim = {0.7cm 0.3cm 2.4cm 0.2cm}, clip, width = 0.5\textwidth]{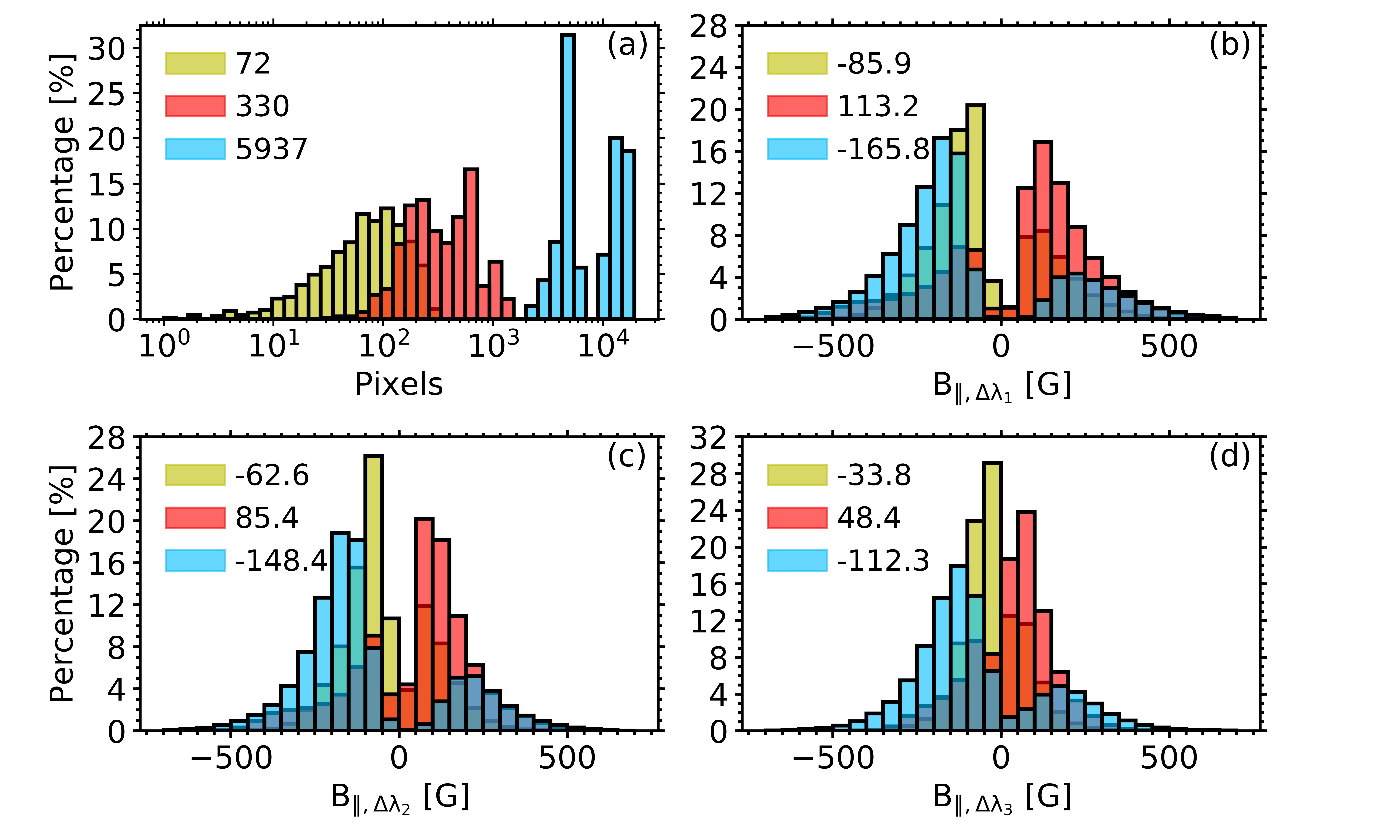}}
\caption{Histograms of the number of pixels with two-lobed Stokes~$V$ signals (panel a), and $B_{\|}$ values for $\mathrm{\Delta \lambda_1}$, $\mathrm{\Delta \lambda_2}$, and $\mathrm{\Delta \lambda_3}$ (panels b--d) in the studied cases. Distributions of pixels within small, medium-sized, and large field concentrations are colored in yellow, red, and blue, respectively. The median value of each distribution is shown in the upper-left corner of the panels.}
\label{fig:hist1}
\end{figure}

We divided the structures we found into three groups: small, medium-sized, and large field concentrations based on their aspect and size at the original cadence. Small field concentrations are elements with lengths of 0.45\arcsec -- 1.2\arcsec wandering around granules. Medium-sized ones seem to be chains of magnetic elements expanding 3.5\arcsec -- 5.5\arcsec \ between granules. Finally, large concentrations are fixed structures spreading over 10\arcsec -- 20\arcsec.

Figure~\ref{fig:hist1} shows histograms of the number of pixels with two-lobed $V$~profiles and $B_{\|}$ estimates for each $\Delta \lambda$ per group, which are normalized to the number of scans and pixels throughout the time sequences in each group ($\sim$96$\times$10$^{3}$, 266$\times$10$^{3}$, and 682$\times$10$^{3}$ pixels for small, medium-sized, and large concentrations), respectively. On average, pixels inside small, medium-sized, and large concentrations represent 0.06\%, 0.13\%, and 2.39\% of the FOV per frame, respectively.

The distribution of the number of pixels with two-lobed $V$~signals in large concentrations diverges from those in small and medium-sized ones. The former has two compact peaks standing for the two cases classified as large concentrations, while the latter two groups show broader size variability. Small concentrations tend to show negative polarity and often have $B_{\|}$ values below 250~G for all $\Delta \lambda$. Medium-sized concentrations usually have positive polarity and harbor estimates below 400~G. The distributions of the large concentrations are smoother and reach $B_\|$ values above 400~G. In addition, most of the pixels in large concentrations have negative polarity, which is the polarity of the largest structure. Generally, all distributions appear shifted to weaker estimates as $\Delta \lambda$ narrows down. 

All cases of each group share similar properties. We therefore describe them by showing one case for each group.

\begin{figure*}[t] 
\centerline{\includegraphics[trim = {0.2cm 0.3cm 0.2cm 0.2cm}, clip, width = 0.95\textwidth]{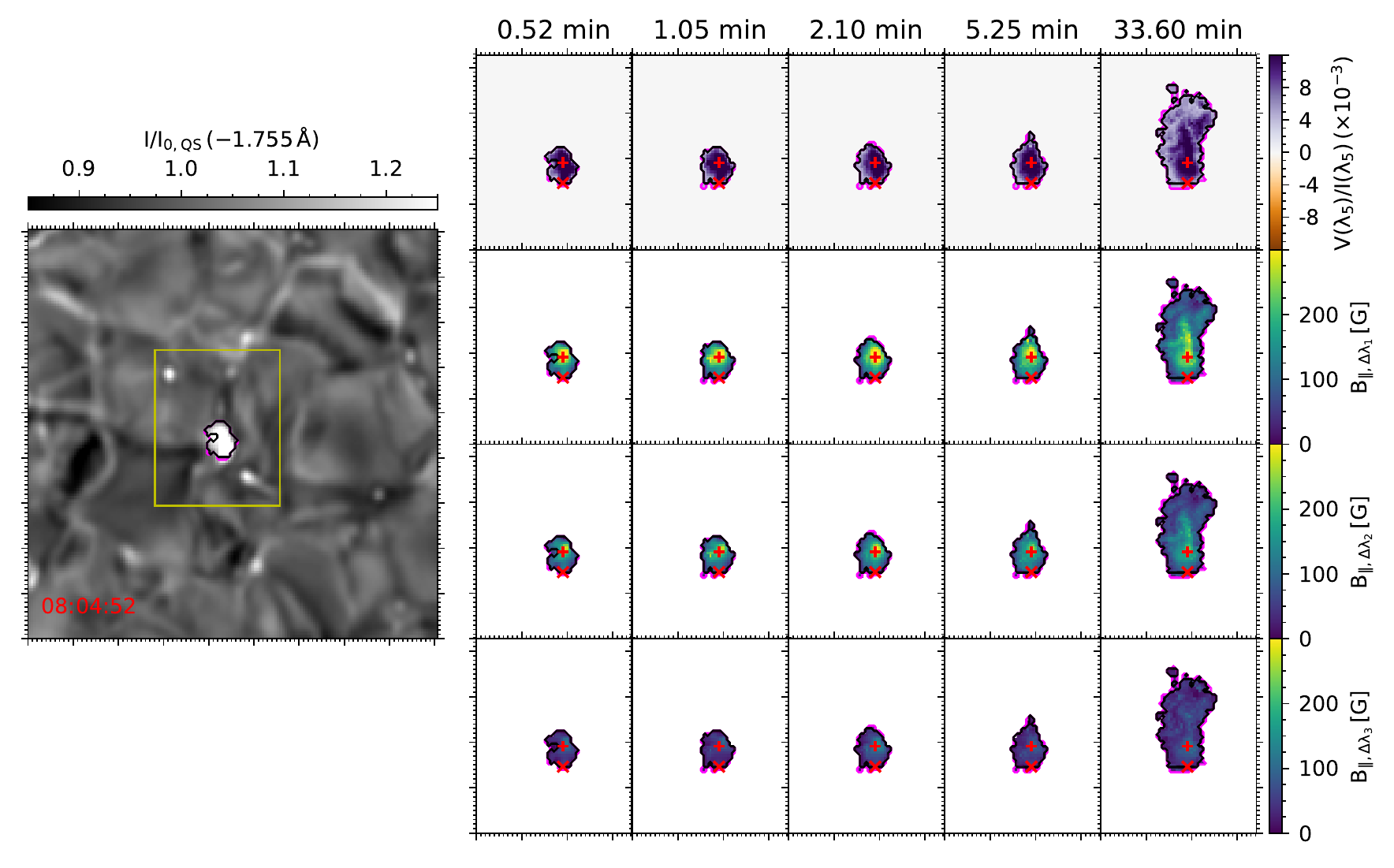}}
\caption{Magnetograms at $-$0.39 \AA\, and $B_{\|}$ estimates on a small field concentration at different $\mathrm{\Delta t}$. Left: \ion{Ca}{II} 854.2~nm intensity filtergram at $-$1.755~\AA. The yellow rectangle encloses the region shown on the right-hand side. Right (from top to bottom): Blue-wing magnetograms at $-$0.39~\AA \, and $B_{\|}$ estimates from $\mathrm{\Delta \lambda_{1}}$, $\mathrm{\Delta \lambda_{2}}$, and $\mathrm{\Delta \lambda_{3}}$. Pink contours delimit the structure, black ones enclose pixels with two-lobed $V$~signals inside it. The plus symbols ($+$) and crosses ($\times$) mark the pixels shown in the two upper and lower rows of Fig.~\ref{fig:gr1_fit}, respectively. Major tick marks represent 1\arcsec. A movie showing the temporal evolution of the parameters displayed in this figure is available online.}
\label{fig:gr1_2}
\end{figure*}

\begin{figure*}[t]
\centerline{\includegraphics[trim = {3.8cm 0.7cm 5.4cm 2cm}, clip, width = 0.95\textwidth]{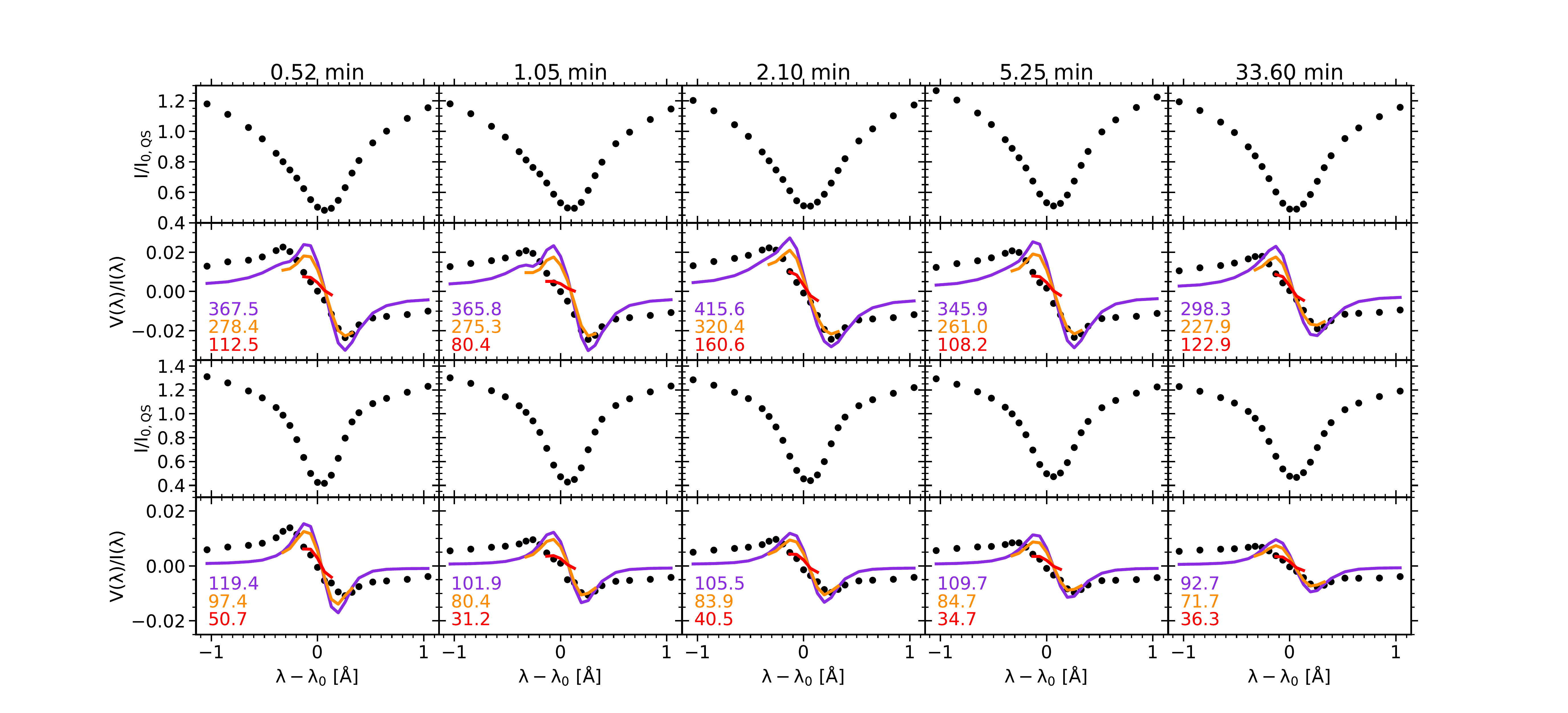}}
\caption{Stokes~$I$ and~$V$ profiles emerging from a small concentration for different $\mathrm{\Delta t}$ (black dots). For visualization purposes, intensity profiles are normalized by the percentage of scans involved in the signal addition. Violet, orange, and red curves stand for the $\delta I(\lambda)/\delta \lambda$ functions scaled by $-C B_{\|}$ for $\mathrm{\Delta \lambda_{1}}$, $\mathrm{\Delta \lambda_{2}}$, and $\mathrm{\Delta \lambda_{3}}$, respectively. $B_{\|}$ values for $\mathrm{\Delta \lambda_{1}}$, $\mathrm{\Delta \lambda_{2}}$, and $\mathrm{\Delta \lambda_{3}}$ in G are shown on each panel in violet, orange, and red.}
\label{fig:gr1_fit}
\end{figure*}

\subsection{Small field concentrations} \label{sub:small}

We identified 19 cases as small concentrations. They appear in intergranular lanes and are highly influenced by neighboring granules. We focus on the case labeled `3' in the first row of Fig.~\ref{fig:data1}. During its temporal evolution (see the movie attached to Fig.~\ref{fig:gr1_2}), small positive-polarity field concentrations appear between granules and move closer to each other until they form a single structure. After that, the structure changes its shape in reaction to the motion of the surrounding granules.

Figure~\ref{fig:gr1_2} also displays the variation of the blue-wing magnetogram (at --0.39~\AA) and $B_{\|}$ with the scanning time $\mathrm{\Delta t}$. At the original cadence, estimates usually range from 100 to 360~G for $\mathrm{\Delta \lambda_{1}}$, and decay to 80--250~G and to 40--150~G for $\mathrm{\Delta \lambda_{2}}$ and $\mathrm{\Delta \lambda_{3}}$. Due to its dynamic behavior, the structure appears larger in maps at long $\mathrm{\Delta t}$. Furthermore, $V$~signals and $B_{\|}$ estimates vary with $\mathrm{\Delta t}$ depending on the individual $I$ and $V$ profiles that are summed and the number of scans involved. Despite pixel-by-pixel differences, the estimates obtained for each $\mathrm{\Delta \lambda}$ do not differ significantly with $\mathrm{\Delta t}$. We remind the reader that $\mathrm{\Delta t}$ refers to the time used to calculate the averaged Stokes $I$ and $V$ profiles (see Sect.~\ref{sec:analysis}).

We assessed the validity of the WFA conditions by comparing the observed Stokes~$V$ profiles to the $I$~derivatives scaled by the estimates for each $\Delta \lambda$. Figure~\ref{fig:gr1_fit} displays these comparisons at two different locations as $\mathrm{\Delta t}$ increases. The two upper and lower rows show the $I$ and $V$ profiles from a pixel at the center and from another closer to the edge of the concentration, respectively. The plus symbols ($+$)  and crosses ($\times$)  in Fig.~\ref{fig:gr1_2} indicate these two specific locations. Although these locations do not share the same features, their $I$ and $V$ signals rely on the imprints left at these pixels during the instants involved in the signal addition. Consequently, all the estimates vary with the scanning time. We note that the gradient between the estimates obtained for each $\mathrm{\Delta \lambda}$  changes very little with $\mathrm{\Delta t}$.

From a qualitative comparison, we observe a systematic divergence between the Stokes~$V$ profiles and the scaled $I$~derivatives at the position of the blue lobe. Both curves only match in some pixels for $\mathrm{\Delta \lambda_{3}}$ (as in the fourth row of Fig.~\ref{fig:gr1_fit}), but they slightly differ in others due to a relative shift between Stokes~$I$ and~$V$ at the line core (see the second row). This relative shift may be due to the coexistence of multiple components in the same resolution element. Using wider spectral ranges, the divergences between the $V$~profiles and the scaled $I$ derivatives remain regardless of $\mathrm{\Delta t}$.

At this point, we note that we only describe the goodness of the fits from a qualitative point of view. A quantitative description is also possible. In that case, we would conclude that the quality worsens with $\mathrm{\Delta t}$ because the noise level for $V$/$I$ decreases with the square root of the number of scans involved in each sum. This is mostly a consequence of the WFA not being able to explain the observations independently of $\mathrm{\Delta t}$. As such a conclusion seems counterintuitive to what we observe from a naked-eye comparison, we opted for a qualitative description of the quality of the fits.

\begin{figure}[t]
\centerline{\includegraphics[trim = {0.2cm 0cm 0.2cm 0cm}, clip, width = 0.5\textwidth]{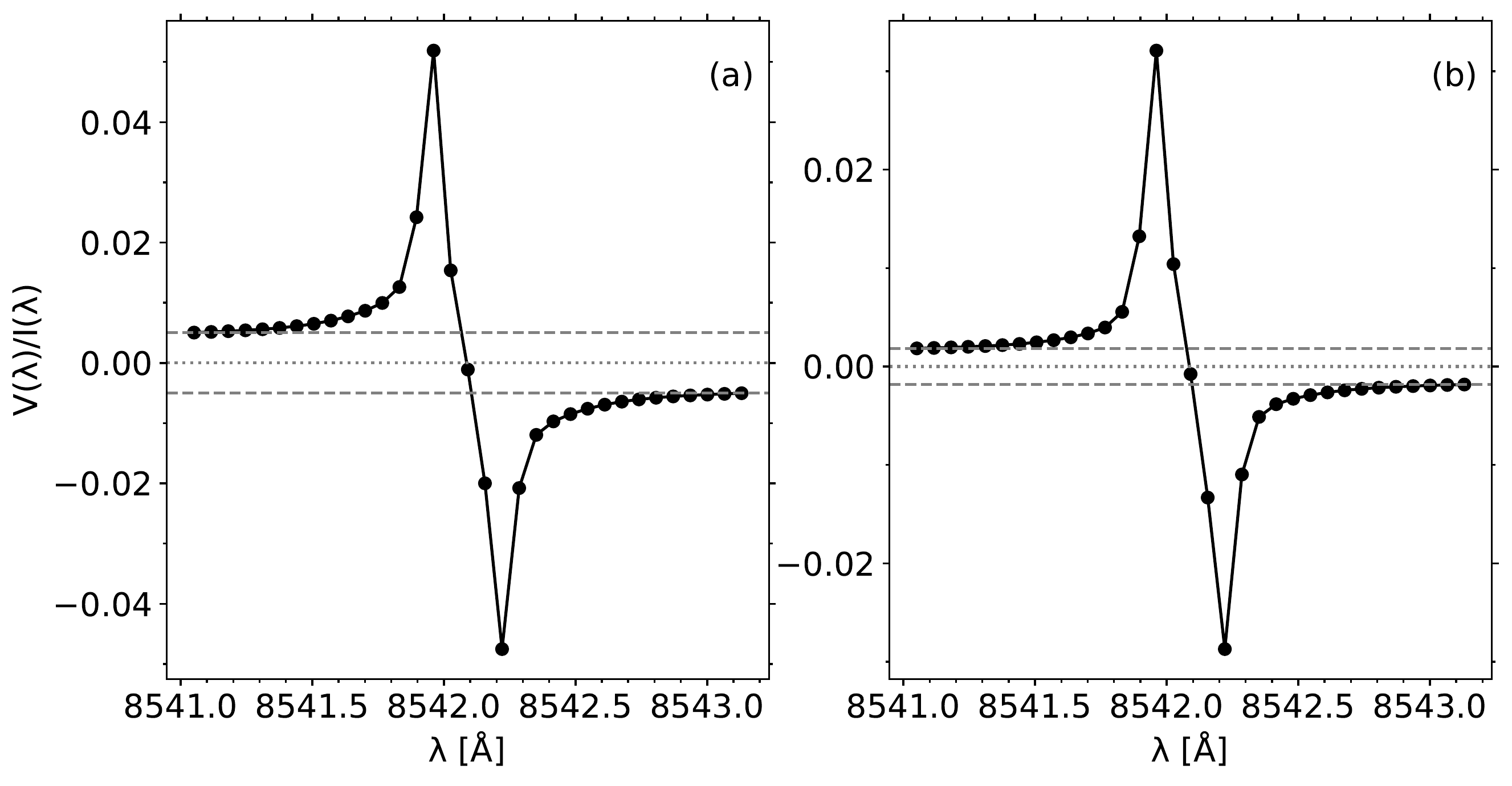}}
\caption{Synthetic $V$ profiles computed from different gradients in $B_{\|}$ between log($\mathrm{\tau_{500}}$) = 1.3 and $-$5.3: from 1\,000 to 100 G (panel a) and from 300 to 100 G (panel b). The dashed gray lines mark the $V$ values at the first and last wavelength positions, and the dotted gray lines indicate zero polarization signal.}
\label{fig:syn}
\end{figure}

In addition, we observe that the fit quality between Stokes~$V$ and the scaled $I$~derivatives at the outer-wing positions improves close to the concentration edge (fourth row in Fig.~\ref{fig:gr1_fit}). To examine this further, we compared the synthetic \ion{Ca}{II} 854.2~nm $V$~profiles resulting from two model atmospheres with the STiC code \citep{2019A&A...623A..74D}. The model atmospheres are based on the FALC model \citep{1993ApJ...406..319F} to which we added $B_{\|}$ values monotonically decreasing between log($\mathrm{\tau_{500}}$) = 1.3 and $-$5.3: (1) from 1\,000 to 100~G, and (2) from 300 to 100~G. We observe similarities in the outer-wing positions of the synthetic $V$ profiles (shown in Fig.~\ref{fig:syn}) and the observed ones, which indicates the high sensitivity of the  $V$ signals of the outer-wings to $B_\|$ in the photosphere. Therefore, the fit quality in the outer-wing positions relies not only on the spectral range but also on the pixel location, as the field concentrations fan out with height. Other features may involve gradients in other parameters or the presence of an unresolved magnetic component not included here for simplicity.

\begin{figure}[!t]
\centerline{\includegraphics[trim = {0cm 0.5cm 1.3cm 1.5cm}, clip, width = 0.45\textwidth]{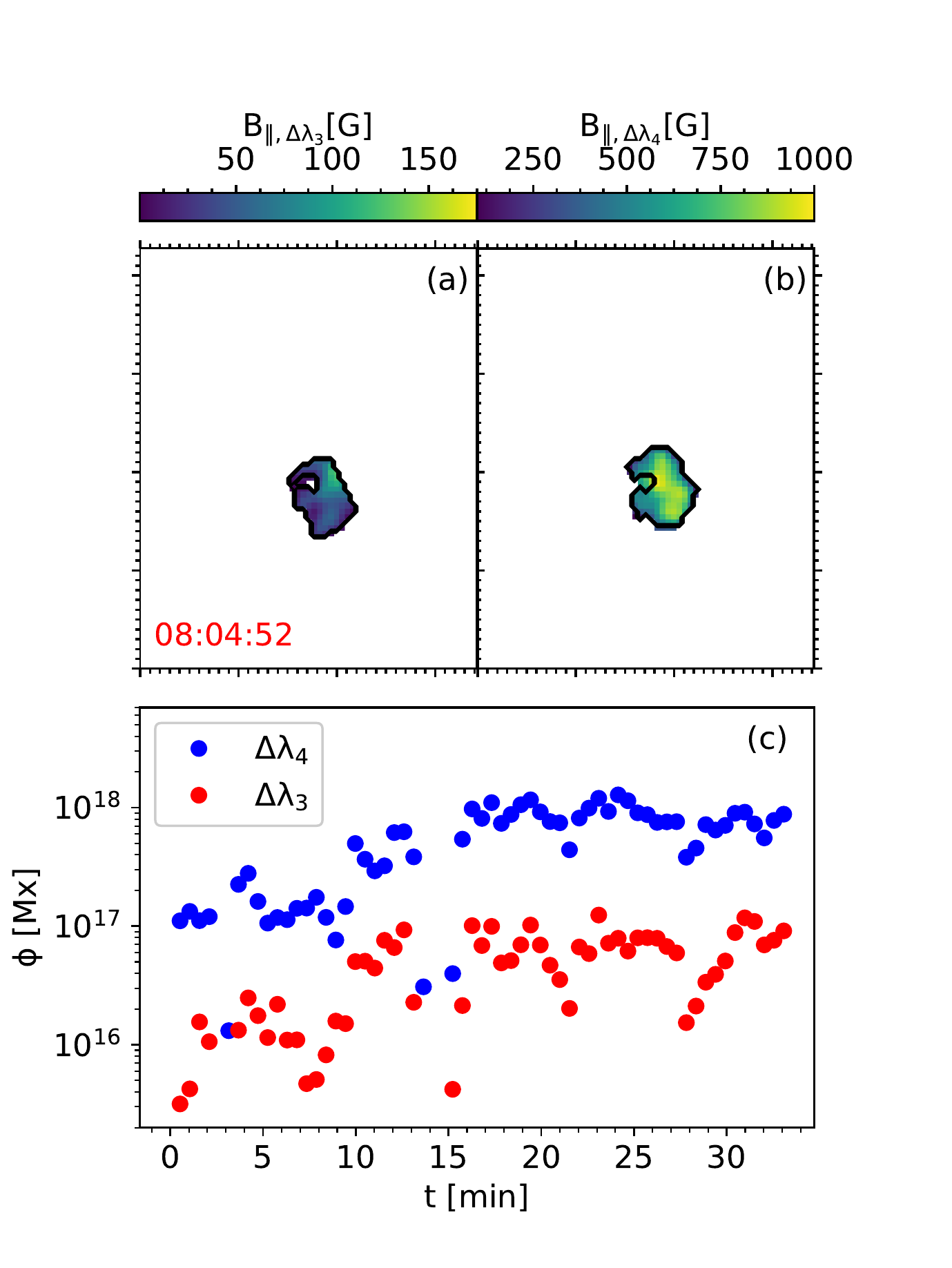}}
\caption{Chromospheric and photospheric estimates in the small concentration at the original cadence (panels a and b), and temporal evolution of the corresponding magnetic flux (panel c). Major tick marks represent 1\arcsec. Black contours enclose pixels with two-lobed $V$~profiles. We note that the values represented in panels (a) and (b) are saturated using different scales, which range from 0 to 175~G and from 100 to 1\,000~G, respectively.}
\label{fig:gr1_flux}
\end{figure}

\begin{figure*}[t] 
\centerline{\includegraphics[trim = {0.25cm 0.3cm 0.3cm 0.2cm}, clip, width = 0.95\textwidth]{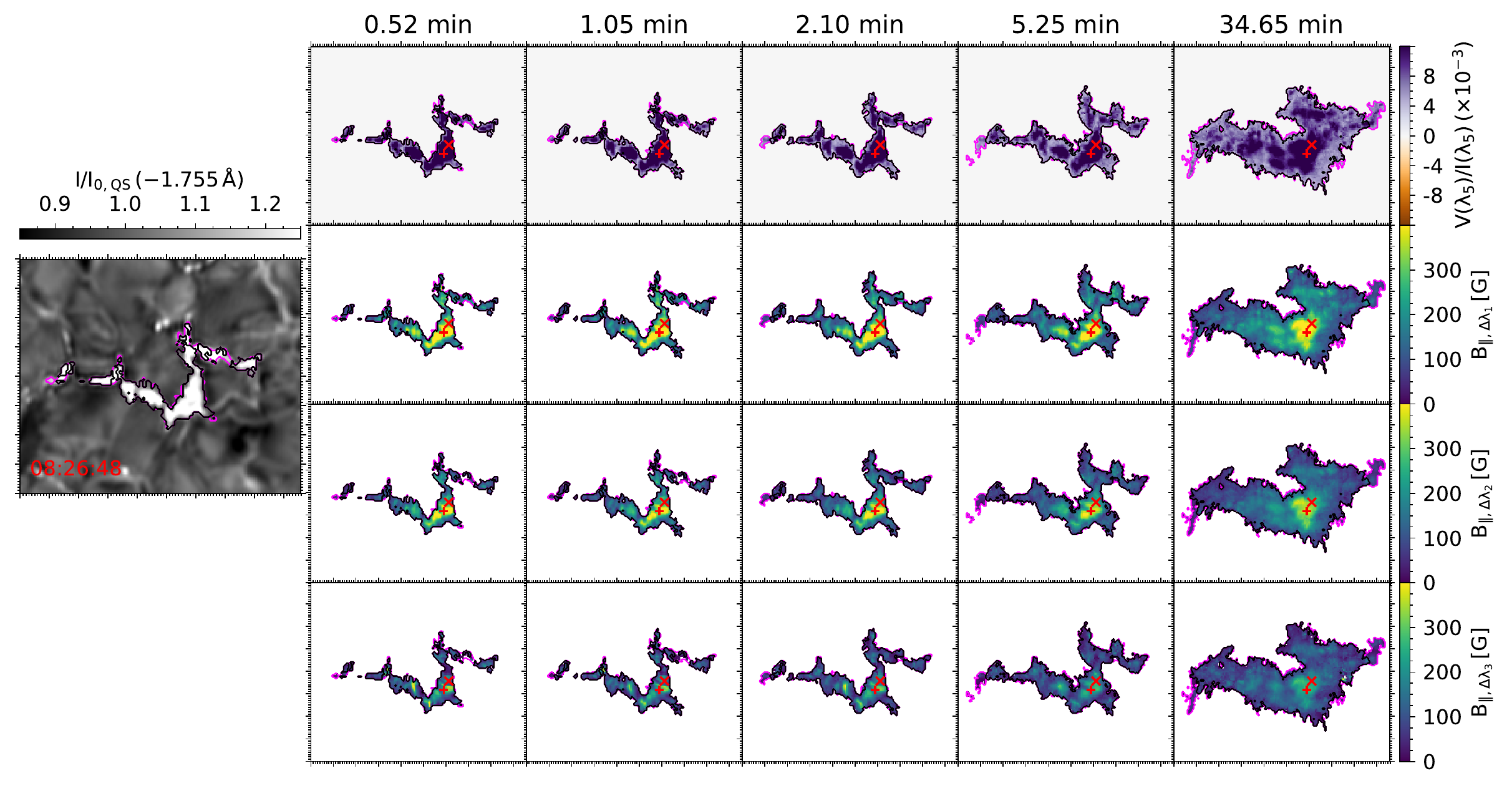}}
\caption{Magnetograms at $-$0.39 \AA\, and $B_{\|}$ maps on a medium-sized concentration for different $\mathrm{\Delta t}$. The plus symbols ($+$)  and crosses ($\times$)  mark the pixels displayed in the two upper and lower rows of Fig.~\ref{fig:gr2_fit}, respectively. Same layout as in Fig.~\ref{fig:gr1_2}. An animated version of this figure displaying the temporal evolution of this case is available online.}
\label{fig:gr2_2}
\end{figure*}

\begin{figure*}[t]
\centerline{\includegraphics[trim = {3.5cm 0.7cm 5.2cm 2cm}, clip, width = 0.95\textwidth]{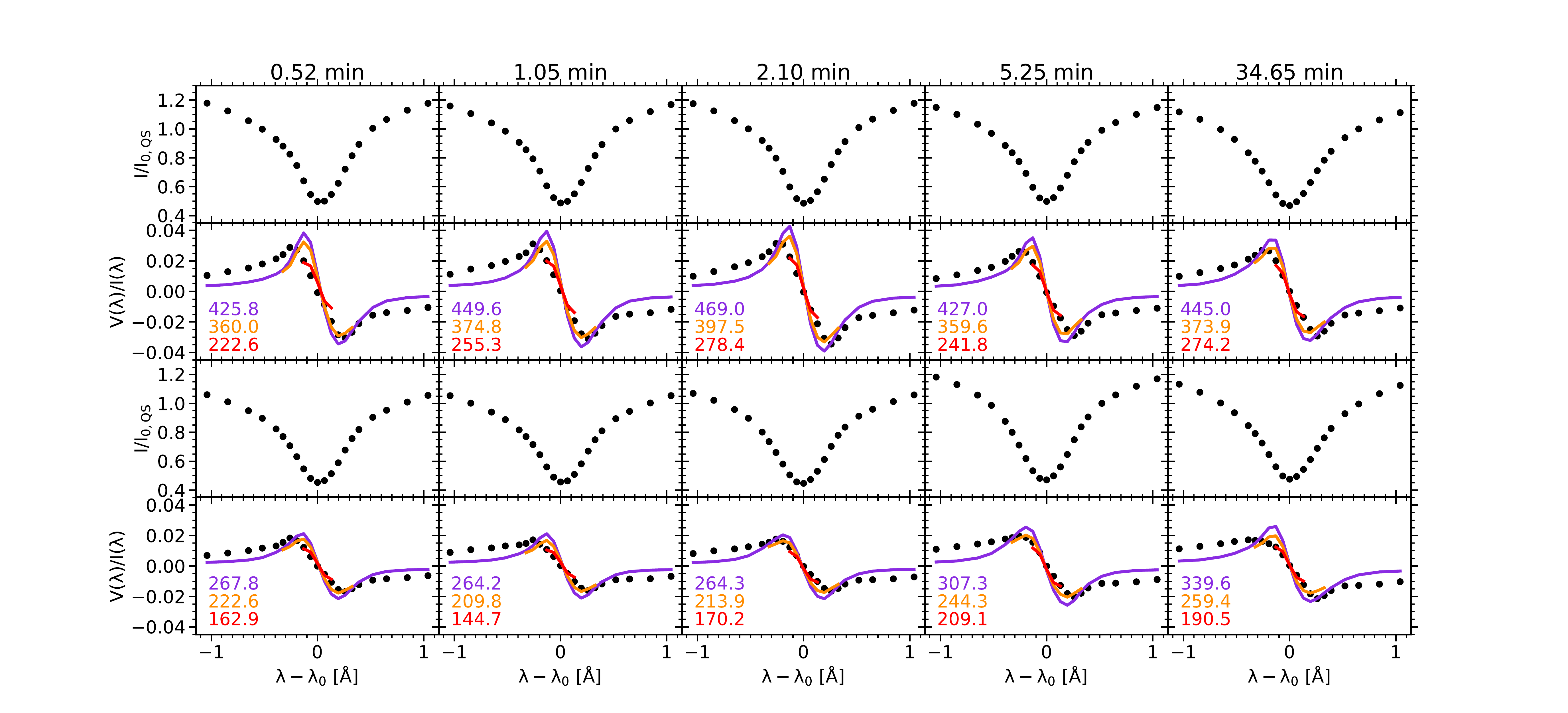}}
\caption{Stokes~$I$ and $V$ profiles coming from the depicted medium-sized concentration for different $\mathrm{\Delta t}$, and their comparison to the Stokes $I$ derivatives scaled by $-C B_{\|}$ for $\mathrm{\Delta \lambda_{1}}$, $\mathrm{\Delta \lambda_{2}}$, and $\mathrm{\Delta \lambda_{3}}$, respectively. Same layout as in Fig.~\ref{fig:gr1_fit}.}
\label{fig:gr2_fit}
\end{figure*}

In view of the above findings, we compared the estimates for $\mathrm{\Delta \lambda_{3}}$ and the outer-wing positions (between $\pm$1.040 and $\pm$0.520~\AA), from which we expect mainly chromospheric and photospheric contributions, respectively. We denote the latter range as $\mathrm{\Delta \lambda_{4}}$. As it may change with height, we delimited the structure again using intensity maps at --0.520~\AA, where we still discern the structure despite the presence of fibrils. However, Fig.~\ref{fig:gr1_flux}a--b shows that the structure barely varies with height, whereas $B_{\|}$ varies conspicuously. Central and outer regions of the concentration show significant height gradients in $B_{\|}$ that resemble those used in the synthesis with STiC. Moreover, these gradients are consistent with time and are independent of the profile smoothing and of the noise decrease, as they appear regardless of $\mathrm{\Delta t}$ (see Fig.~\ref{fig:gr1_fit}). Therefore, WFA conditions cannot be satisfied by adding signal.

Finally, we computed the magnetic flux from the chromospheric and photospheric contributions using the expression

\begin{equation}
\Phi(B_{\|}) \ = \ \sum_{i,j}B_{\|}(i, j) \cdot \Delta A,
\label{eq:flux}
\end{equation}

\noindent where $B_{\|}(i, j)$ is the estimate for $\mathrm{\Delta \lambda_{3}}$ and $\mathrm{\Delta \lambda_{4}}$ at a pixel with coordinates $(i, j),$ and $\Delta A$ is the pixel area. Specifically, the $B_{\|}$ values used in Eq.~(\ref{eq:flux}) are those estimated at the original cadence. Figure~\ref{fig:gr1_flux}c shows the resulting magnetic flux values, which vary with time depending on the temporal evolution of the structure. In addition, given the height gradient in $B_\|$, the photospheric and chromospheric magnetic flux values differ by one order of magnitude during the time interval showing this case (see also Table~\ref{tab:flux}).

\begin{table}[t]
\caption{Statistics on the chromospheric and photospheric magnetic flux in the depicted field concentrations}
\begin{tabular}{l c c c c}
\hline \hline
Field & \multicolumn{2}{c}{$\Phi (B_{\|,  \ \mathrm{\Delta \lambda_{3}}})$} & \multicolumn{2}{c}{$\Phi (B_{\|,  \ \mathrm{\Delta \lambda_{4}}})$} \\
Concent. & Median & $\sigma$ & Median & $\sigma$ \\
 & (Mx) & (Mx) & (Mx) & (Mx) \\
\hline
Small & 4.7$\times$10$^{16}$ & 3.5$\times$10$^{16}$& 5.6$\times$10$^{17}$&3.7$\times$10$^{17}$ \\ 
Medium & 1.7$\times$10$^{18}$ & 3.2$\times$10$^{17}$ & 1.0$\times$10$^{19}$&1.4$\times$10$^{18}$\\ 
Large & $-$2.9$\times$10$^{19}$ & 2.5$\times$10$^{19}$ & $-$1.1$\times$10$^{20}$ & 6.3$\times$10$^{18}$ \\  
\hline
\end{tabular}
\label{tab:flux}
\end{table}

\subsection{Medium-sized field concentrations} \label{sub:gr2}

We classified ten cases as medium-sized field concentrations; they have a practically fixed position and are observed as stable structures. In this subsection, we show an analysis of the case labeled `3' in the second series (see Fig.~\ref{fig:data1}), which is portrayed in Fig.~\ref{fig:gr2_2}. This positive-polarity case undergoes fragmentations and mergings (see the animation attached to Fig.~\ref{fig:gr2_2}). At the original cadence, the $B_{\|}$ estimates for $\mathrm{\Delta \lambda_{1}}$ and $\mathrm{\Delta \lambda_{2}}$ are usually above 200~G and exceed 350~G at the center of the structure. These values are weaker for $\mathrm{\Delta \lambda_{3}}$ overall, reaching $\sim$250--300~G. Because of its stable behavior, $B_{\|}$ maps barely change for $\mathrm{\Delta t}$ $<$ 5.25 minutes.

The intensity profiles at the center of the concentration are different from those emerging closer to the edges, as shown in the upper and lower rows of Fig.~\ref{fig:gr2_fit}. Locations close to the edge usually show line asymmetries and shifts, which may indicate that the physical conditions at the center of the structure are firmer than in the outer regions. In particular, the intensity profiles represented in the lower panels display slight asymmetries compared to those in the upper rows. Stokes~$V$ profiles also differ as amplitudes weaken away from the center. Aside from the $V$ amplitude variations, we do not find changes with $\mathrm{\Delta t}$ worth mentioning. Regarding the $B_{\|}$ estimates, the central and outer locations of the concentration show values of $\sim$420 and 260~G for $\mathrm{\Delta \lambda_{1}}$ at the original cadence, respectively. These $B_{\|}$ weaken by about 60 G and 100--200~G for $\mathrm{\Delta \lambda_{2}}$ and $\mathrm{\Delta \lambda_{3}}$, respectively. Moreover, the gradient between the estimates for each spectral range is stable with $\mathrm{\Delta t}$ and comparable to the individual values.

\begin{figure}[t]
\centerline{\includegraphics[trim = {0.3cm 0cm 0.4cm 0.2cm}, clip, width = 0.43\textwidth]{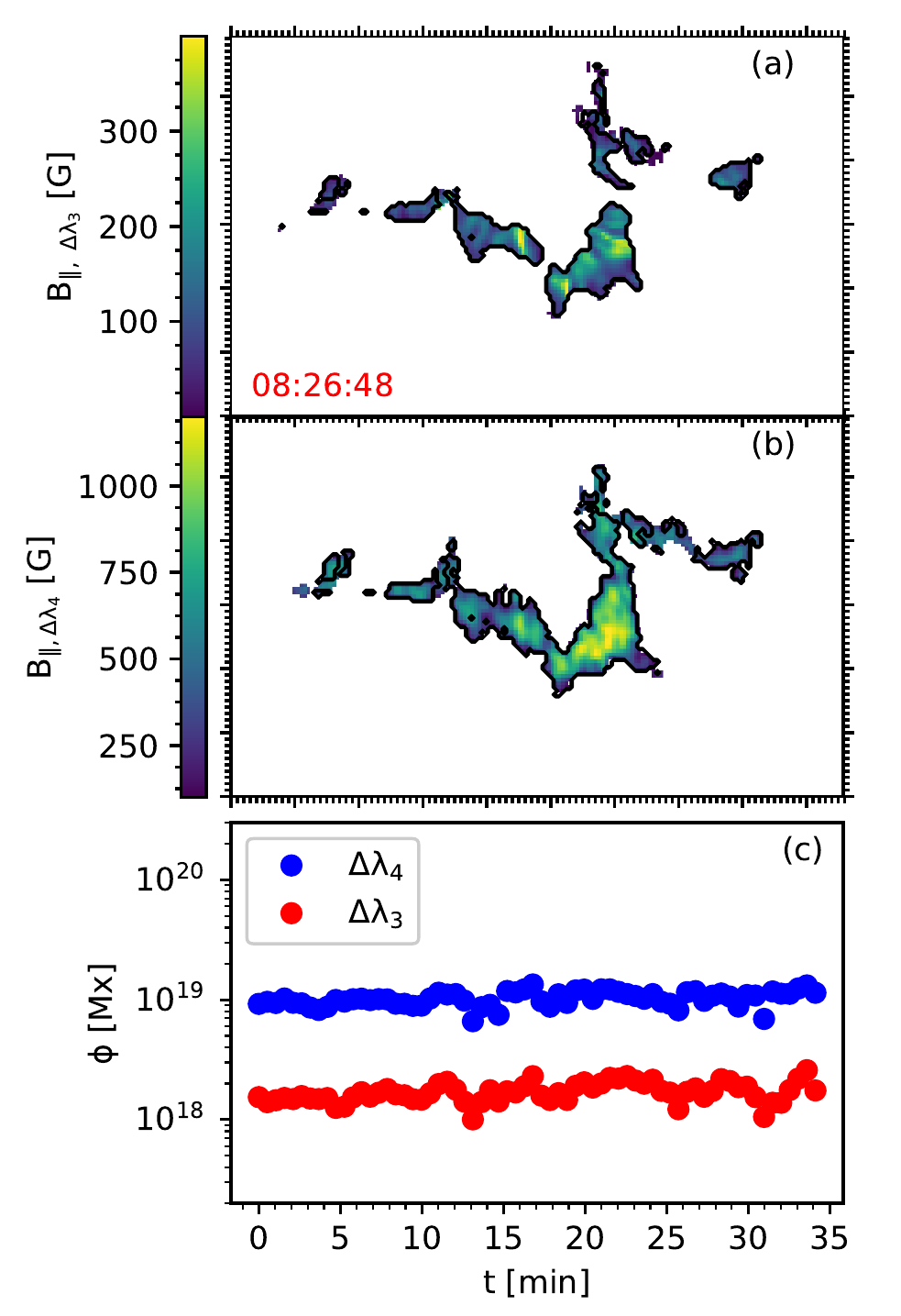}}
\caption{Chromospheric and photospheric $B_{\|}$ estimates in the medium-sized case at the original cadence (panels a and b), and temporal evolution of the corresponding magnetic flux (panel c). Major tick marks represent 1\arcsec. Black contours enclose pixels with two-lobed $V$~profiles. We note that the values represented in panels (a) and (b) are saturated using different scales, which range from 0 to 400~G and from 100 to 1\,200~G, respectively.}
\label{fig:gr2_flux}
\end{figure}

We also compared the mostly chromospheric and photospheric $B_{\|}$ estimates in this structure. The inferred values diverge significantly (see Fig.~\ref{fig:gr2_flux}a--b), having median values of 105 and 524~G, respectively. Thus, the magnetic flux values for $\mathrm{\Delta \lambda_{3}}$ and $\mathrm{\Delta \lambda_{4}}$ differ again by one order of magnitude (Fig.~\ref{fig:gr2_flux}c and Table~\ref{tab:flux}) but, unlike the small case, they are stable with time. The height gradient in $B_{\|}$ also explains the poor fit quality between the $V$~profiles and the scaled $I$ derivatives for $\mathrm{\Delta \lambda_{1}}$ and $\mathrm{\Delta \lambda_{2}}$. In contrast, fits are usually better for $\mathrm{\Delta \lambda_{3}}$, although they are affected by relative shifts between Stokes $I$ and $V$ at some positions. All in all, the estimates for wider spectral ranges are unsafe as the WFA conditions are unfulfilled.

\subsection{Large field concentrations} \label{sub:gr3}

\begin{figure*}[t] 
\centerline{\includegraphics[trim = {0.3cm 0.3cm 0.2cm 0.2cm}, clip, width = 0.93\textwidth]{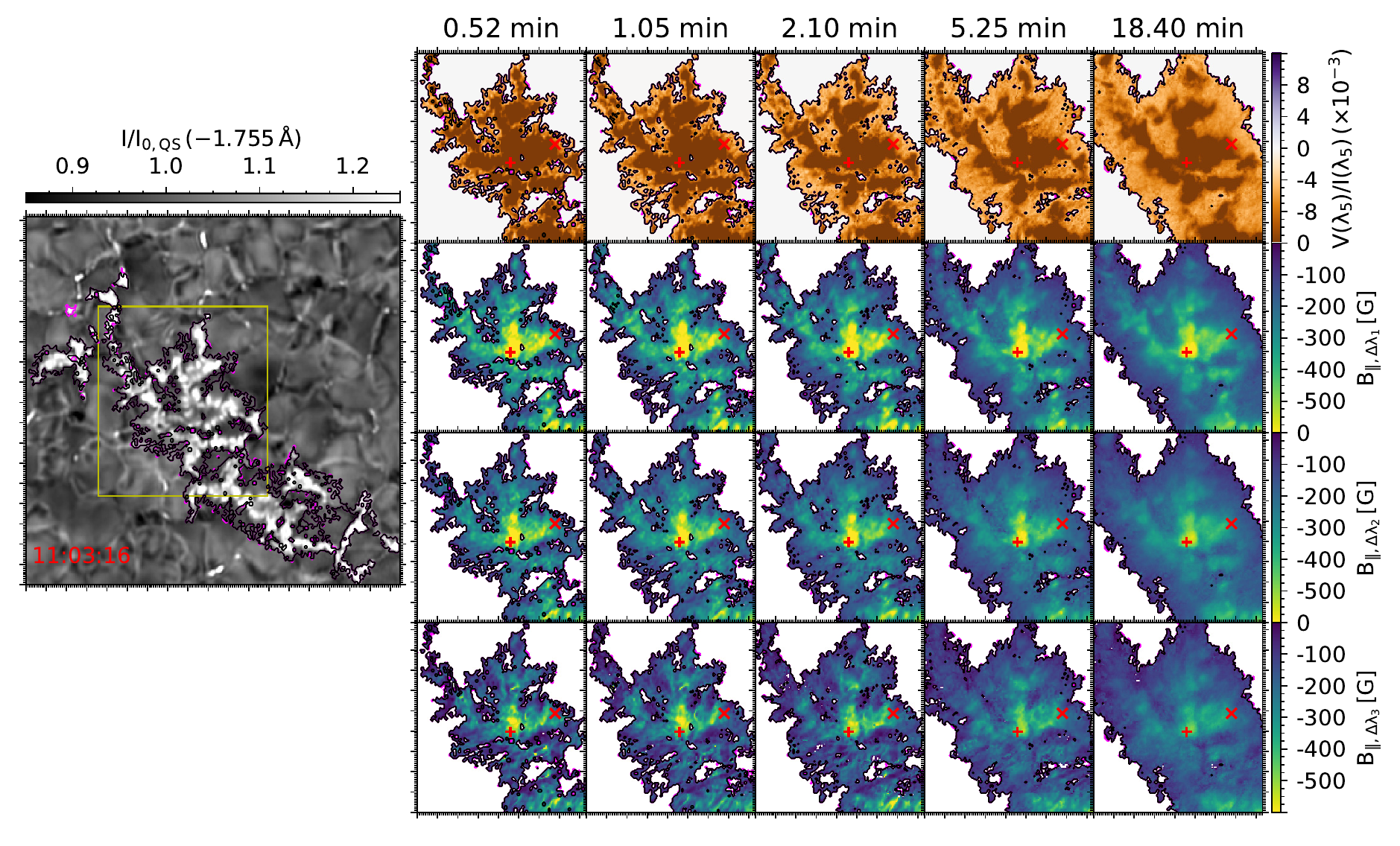}}
\caption{Magnetograms at $-$0.39 \AA\, and $B_{\|}$ estimates on a large concentration for different $\mathrm{\Delta t}$. The plus symbols ($+$) 
and crosses ($\times$)  mark the pixels shown in the two upper and lower rows of Fig.~\ref{fig:gr3_fit}, respectively. The yellow rectangle delimits the region displayed on the right-hand side. Same layout as in Fig.~\ref{fig:gr1_2}. A movie showing the temporal evolution of the parameters displayed in this figure is available online.}
\label{fig:gr3_2}
\end{figure*}

We identified two cases of large field concentrations; they are labeled `1' and `2' in the last row of Fig.~\ref{fig:data1} and may be an example of \textit{enhanced network}. We focus on case 2, shown in Fig.~\ref{fig:gr3_2}, the polarity of which is negative.

The movie attached to Fig.~\ref{fig:gr3_2} shows that the $B_{\|}$ values seem to oscillate with time, mostly at the center of the structure. As these oscillations coincide with overall intensity fluctuations, we think they are related to seeing instabilities instead of actual changes in the structure. At the original cadence, we infer estimates that progressively increase from the edges to the center of the structure, where they are of about $-$750~G for $\mathrm{\Delta \lambda_{1}}$. The $B_{\|}$ maps for $\mathrm{\Delta \lambda_{2}}$ and $\mathrm{\Delta \lambda_{3}}$ are similar to that obtained for  $\mathrm{\Delta \lambda_{1}}$, though estimates are weaker reaching up to $-$650 and $-$550~G, respectively. In addition, the concentration grows with $\mathrm{\Delta t}$ due to interactions with nearby structures and to the variable seeing conditions, whereas $B_{\|}$ values weaken globally.

\begin{figure*}[t]
\centerline{\includegraphics[trim = {3.8cm 0.7cm 5.2cm 2cm}, clip, width = 0.95\textwidth]{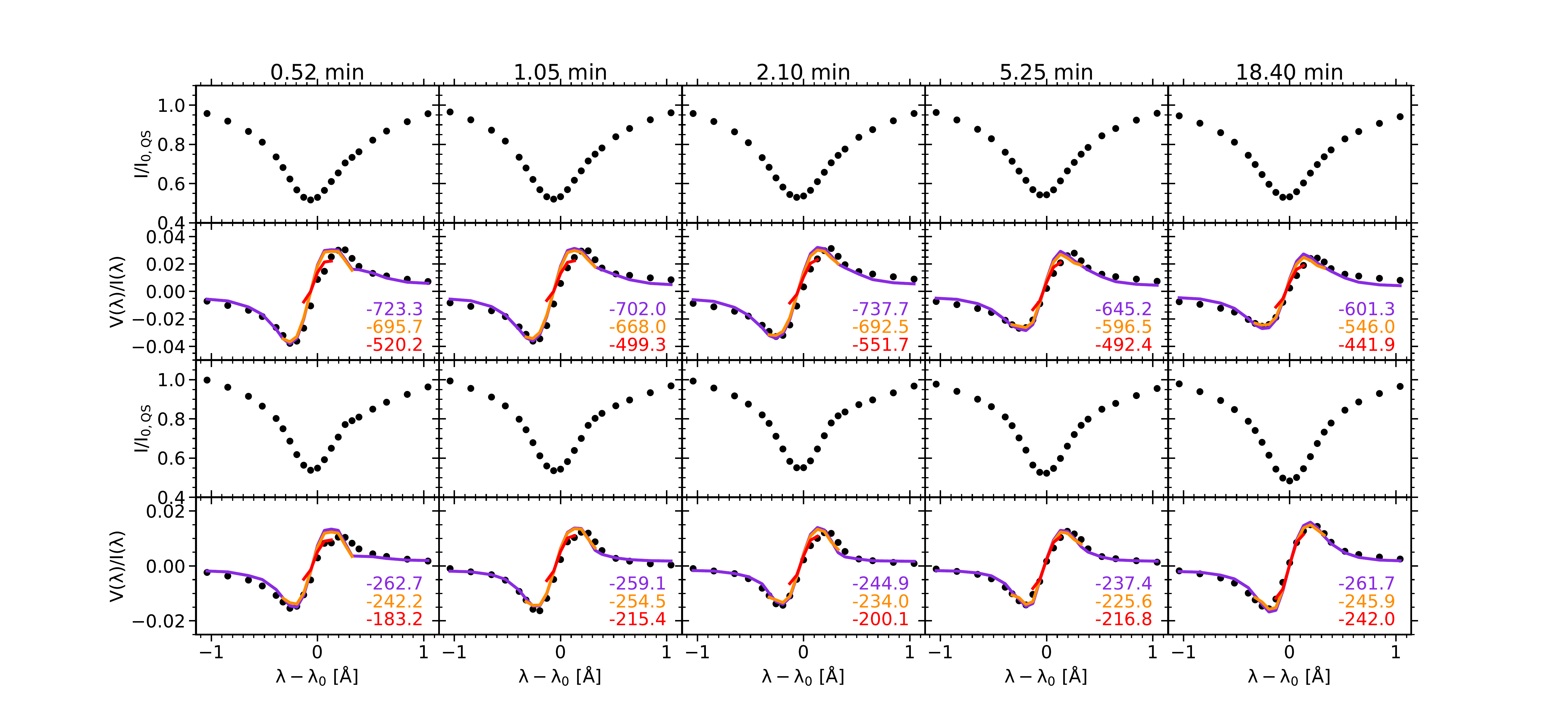}}
\caption{Stokes $I$ and $V$ profiles coming from the depicted large concentration at different $\mathrm{\Delta t}$, and their comparison to the Stokes $I$ derivatives scaled by $-C B_{\|}$ for $\mathrm{\Delta \lambda_{1}}$, $\mathrm{\Delta \lambda_{2}}$, and $\mathrm{\Delta \lambda_{3}}$, respectively. Same layout as in Fig.~\ref{fig:gr1_fit}.}
\label{fig:gr3_fit}
\end{figure*}

At the original cadence, asymmetries and blueshifts are apparent in the intensity profiles from the center of this concentration and from a pixel closer to the edge (see upper and lower rows of Fig.~\ref{fig:gr3_fit}, respectively). These features persist with the scanning time until they soften at the longest $\mathrm{\Delta t}$ due to the profile smoothing. At the same time, $V$~amplitudes decrease with the distance to the center of the structure and barely change with $\mathrm{\Delta t}$. As a result of these changes along the concentration, $B_{\|}$ values differ significantly depending on the pixel location. We find that, at the original cadence, estimates range from $-$500 to $-$700~G and from $-$180 to $-$260~G for the different $\mathrm{\Delta \lambda}$ at the center and closer to the edge, respectively. These $B_\|$ gradients are stable with $\mathrm{\Delta t}$, and so the conditions along the formation region related to the used spectral ranges hardly vary with time.

Although not displayed, the difference between the mostly chromospheric and photospheric estimates in this concentration is not as significant as in smaller structures. The median values in this case for $\mathrm{\Delta \lambda_{3}}$ and $\mathrm{\Delta \lambda_4}$ are $-$160.7 and $-$360.3~G, respectively. Consequently, the photospheric magnetic flux is three to four times greater than the chromospheric flux (see Table~\ref{tab:flux}). Similarly to the medium-sized case, the magnetic flux values inferred in this concentration have a stable temporal evolution. The reduced difference between the photospheric and chromospheric estimates may explain a better match between Stokes~$V$ and the scaled $I$ derivatives regardless of the spectral range, scanning time, and location.

\section{Discussion} \label{sec:discussion}

Assuming WFA conditions, the reliability of the estimates for wide spectral ranges seems to rely on the concentration size. Small and medium-sized field concentrations have steeper time-independent height gradients in $B_{\|}$ compared to their chromospheric $B_{\|}$ values than the large ones, which leads to unreliable estimates of the former as the WFA requisites are unfulfilled. On the contrary, estimates for inner-core ranges are usually reliable in all concentrations.

Despite its assets, some observers may employ other restoration methods on the basis of the liabilities of MOMFBD. In this section, we compare results from data restored with different methods in order to decipher whether or not the restoration technique is related to the fulfillment of WFA conditions. In addition, as the WFA is widely used, we also discuss our results by comparing them to previous studies that used similar data.

\begin{figure*} 
\centerline{\includegraphics[trim = {0.5cm 0.3cm 0.1cm 0.2cm}, clip, height = 0.95\textheight]{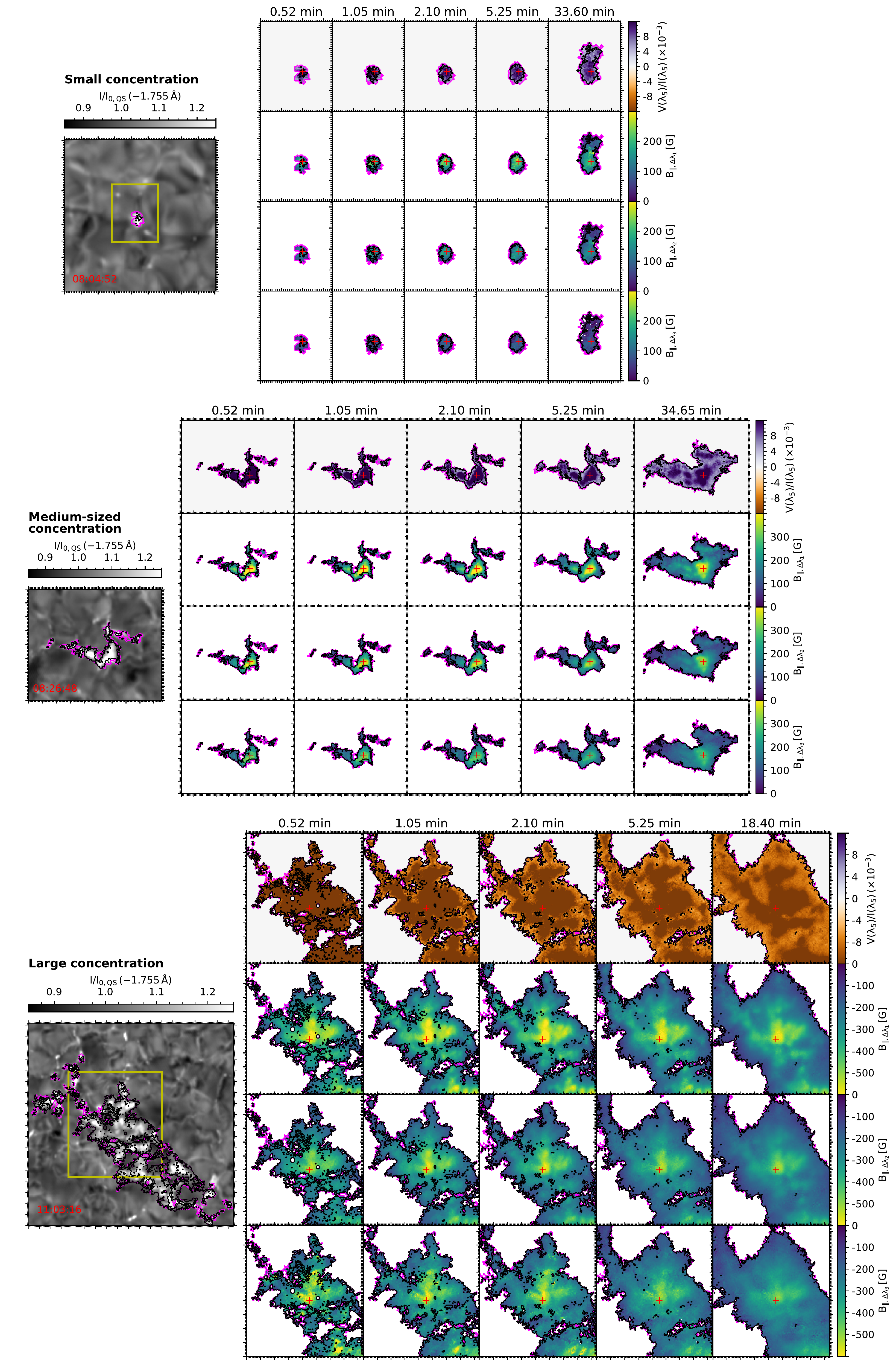}}
\caption{Details of the cases shown in Figs.~\ref{fig:gr1_2},~\ref{fig:gr2_2}, and~\ref{fig:gr3_2} using the NM datasets. Left: \ion{Ca}{II} 854.2~nm intensity filtergrams at $-$1.755~\AA. The yellow rectangles superimposed over the small and large concentrations enclose the regions shown zoomed-in on the right-hand side. Right: Magnetograms at $-$0.39~\AA \, and $B_{\|}$ estimates for each $\Delta \lambda$ and $\mathrm{\Delta t}$. The plus symbols ($\mathrm{+}$) mark the pixels shown in Fig.~\ref{fig:gr123_nm_fit}. Pink and black contours enclose each case and the pixels with two-lobed $V$ profiles inside them, respectively. Major tick marks represent 1\arcsec.}
\label{fig:gr123_nm}
\end{figure*}

\subsection{Results from differently restored data}

MOMFBD is a successful image-correction technique. However, its application comes with high computational cost and may lead to amplified noise levels \citep{2011A&A...533A..21P}.

Alternatively, images can be restored by adding the accumulations acquired per wavelength position. In this case, most of the reduction process is carried out using the CRISPRED pipeline \citep{2015A&A...573A..40D} and the effects of seeing in observations are partially corrected by applying the destretching technique. During this process, differential motions in small subfields of the wide-band images, which are previously derotated and aligned, are measured using cross-correlation. After combining accumulations, the polarimetric calibration is performed as in \citet{2008A&A...489..429V}. Finally, each pair of images with orthogonal polarization states, recorded simultaneously by the narrow-band cameras, are combined to remove the seeing-induced polarization, and the residual crosstalk is corrected. We also corrected each monochromatic polarization image for high-frequency polarimetric interference fringes in a similar way to \citet{2020A&A...644A..43P}. As each image is apodized in this process, we clipped the blurred edges of the output polarization maps, obtaining smaller images. The noise level of the resulting Stokes~$V$ images is in the range of 1.8--2.5$\times$10$^{-3}$ measured on the $V$/$I_{0}$ maps at the first observed wavelength for each series. We refer to the datasets restored with and without the MOMFBD technique as M and NM datasets, respectively.

We examined 27 of the cases shown in Fig.~\ref{fig:data1} because some of them are partially or completely beyond the image boundaries of the NM datasets. Each case has been delimited using thresholds of intensity at $-$1.755~\AA \, and $V(\lambda)/I(\lambda)$ at $-$0.39~\AA\, of 0.95~$I_{0,QS}$ and 4$\times$10$^{-3}$, respectively. After that, we estimated $B_{\|}$ by applying the WFA in pixels with two-lobed $V$ profiles and analyzed their variation with the scanning time, $\mathrm{\Delta t}$ (see Sect.~\ref{sec:analysis}). 

\begin{figure*}[t]
\centerline{\includegraphics[trim = {0cm 1.3cm 2.7cm 1.2cm}, clip, width = 0.95\textwidth]{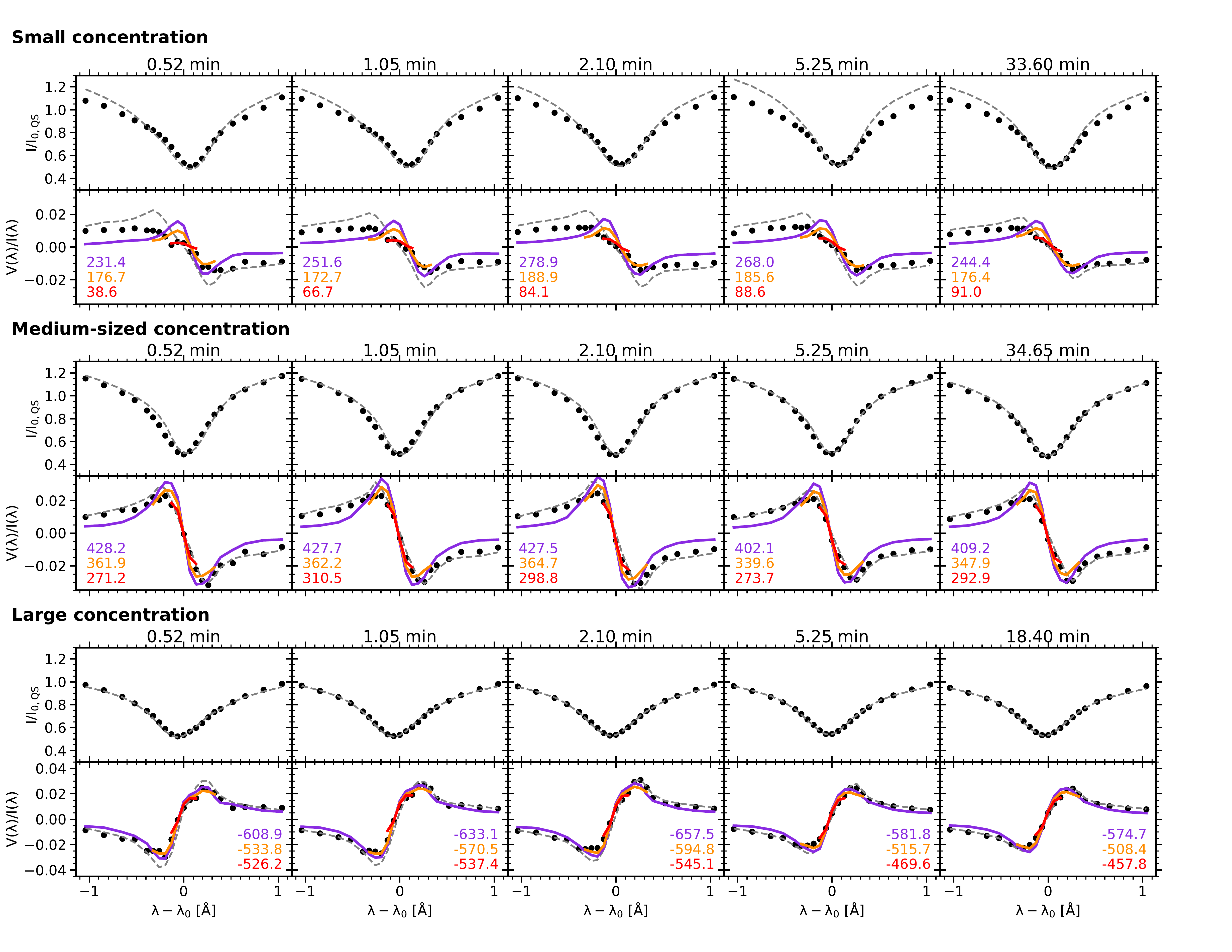}}
\caption{Profiles coming from the pixels marked in Fig.~\ref{fig:gr123_nm} (black dots). For visualization purposes, $I$ profiles are normalized by the percentage of scans involved in the signal addition. Curves in violet, orange, and red stand for the $\delta I(\lambda)/\delta \lambda$ scaled by $-C B_{\|}$ for $\mathrm{\Delta \lambda_{1}}$, $\mathrm{\Delta \lambda_{2}}$, and $\mathrm{\Delta \lambda_{3}}$, respectively. The corresponding $B_{\|}$ values are indicated in G following the same color scheme. Profiles from the M datasets are shown by dashed gray lines.}
\label{fig:gr123_nm_fit}
\end{figure*}

Figure~\ref{fig:gr123_nm} shows the $B_{\|}$ estimates obtained for the cases depicted in Sect.~\ref{sec:analysis} using the NM datasets. The physical quantities in this figure appear smeared in comparison to those in Figs.~\ref{fig:gr1_2}, \ref{fig:gr2_2}, and \ref{fig:gr3_2}. We observe not only blurrier intensity images but also smoother $B_\|$ estimates. Indeed, the latter are generally 50--100~G weaker for all $\mathrm{\Delta t}$. These differences between both datasets appear because the intensity signal in each pixel is spread out over its surroundings in the NM datasets, as neither the blurring effect of the telescope point spread function (PSF) nor the low-altitude seeing is corrected. Nonetheless, the smearing relies on the concentration size. This correlation is clear in Fig.~\ref{fig:gr123_nm_fit}, where the divergence of the $I$ and $V$ profiles from the NM and M datasets increases as the concentration size decreases.

Furthermore, in Fig.~\ref{fig:gr123_nm_fit}, we observe that the fit quality between the Stokes $V$ profiles and the scaled $I$ derivatives from the NM datasets is similar to that described for the M datasets. However, the smearing in the NM datasets leads to better fits in all concentrations, particularly for $\mathrm{\Delta \lambda_3}$. In the case of wider spectral ranges, the fit quality again relies on the concentration size, because height gradients in $B_\|$ are more abrupt in small and medium-sized concentrations. Magnetic flux variations between the photosphere and chromosphere are therefore of the same order as in the M datasets (Table~\ref{tab:fluxnm}). Specifically, the median values of the estimates for the photospheric $\mathrm{\Delta \lambda_4}$ range (and the chromospheric $\mathrm{\Delta \lambda_3}$ one) in the small, medium-sized, and large concentrations are 461.3 (33.7), 475.3 (174.3), and $-$404.9 ($-$248.2)~G, respectively.

\begin{table}[t]
\caption{Statistics on the chromospheric and photospheric magnetic flux in the depicted field concentrations using the NM datasets.}
\begin{tabular}{l c c c c}
\hline \hline
Field & \multicolumn{2}{c}{$\Phi (B_{\|, \ \mathrm{\Delta \lambda_{3}}})$} & \multicolumn{2}{c}{$\Phi (B_{\|, \ \mathrm{\Delta \lambda_{4}}})$} \\
Concent. & Median & $\sigma$ & Median & $\sigma$ \\
 & (Mx) & (Mx) & (Mx) & (Mx) \\
\hline
Small & 1.6$\times$10$^{16}$ & 2.4$\times$10$^{16}$& 2.1$\times$10$^{17}$&1.6$\times$10$^{17}$ \\ 
Medium & 2.0$\times$10$^{18}$ & 3.5$\times$10$^{17}$ & 8.0$\times$10$^{18}$&1.2$\times$10$^{18}$\\ 
Large & $-$2.6$\times$10$^{19}$ & 8.1$\times$10$^{16}$ & $-$8.4$\times$10$^{19}$ & 2.6$\times$10$^{19}$ \\  
\hline
\end{tabular}
\label{tab:fluxnm}
\end{table}

\begin{figure} 
\centerline{\includegraphics[trim = {0.7cm 0.3cm 2.4cm 0.2cm}, clip, width = 0.49\textwidth]{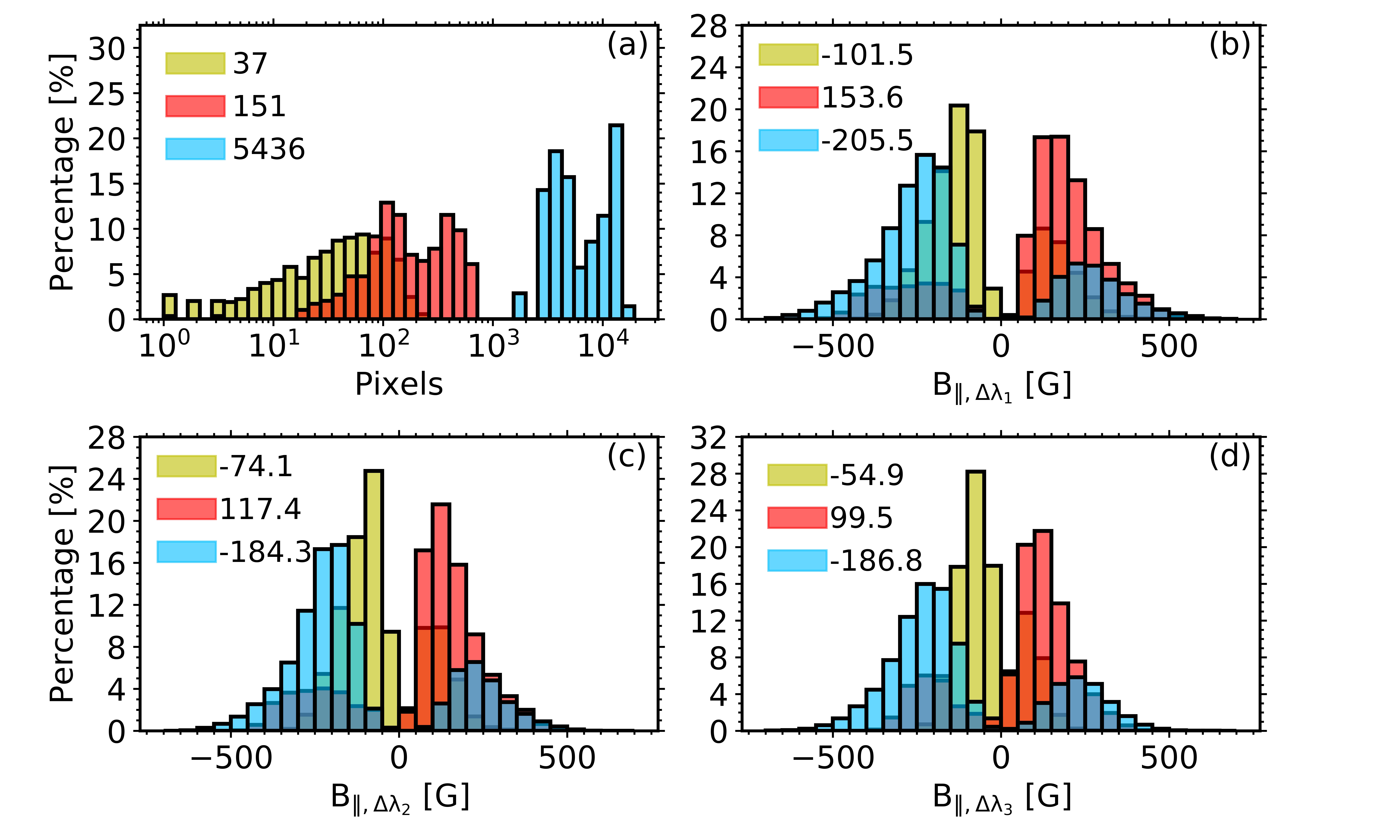}}
\caption{Histograms of the number of pixels with two-lobed Stokes $V$ signals (panel a) and $B_{\|}$ values for $\mathrm{\Delta \lambda_1}$, $\mathrm{\Delta \lambda_2}$, and $\mathrm{\Delta \lambda_3}$ (panels b--d) in the studied cases using the NM datasets. Yellow, red, and blue distributions represent pixels in small, medium-sized, and large field concentrations, respectively. The upper-left corner of the panels show the median value of each distribution.}
\label{fig:hist2}
\end{figure}

Finally, Fig.~\ref{fig:hist2} shows the histograms of the number of pixels with two-lobed $V$ profiles in the concentrations analyzed using the NM datasets and their $B_{\|}$ values for each spectral range. The former distributions are normalized to the number of scans showing each group whereas the latter are normalized to the number of pixels per group in all sequences ($\sim$46$\times$10$^{3}$, 150$\times$10$^{3}$, and 510$\times$10$^{3}$ pixels for small, medium-sized, and large structures), respectively. Although all distributions resemble those for the M datasets (Fig.~\ref{fig:hist1}), they are affected by the value smearing, as fewer pixels meet the threshold used to delimit each case (Fig.~\ref{fig:hist2}a) and the peaks of some distributions are shifted to stronger estimates (Fig.~\ref{fig:hist2}b--d).

\subsection{Comparison with previous studies}

Over the past few years, investigations based on high-spatial-resolution data have reported a weakening of the quiet-Sun magnetic fields with height (e.g., \citealt{2018ApJ...857...48G}; \citealt{2019A&A...621A...1R}; \citealt{2020A&A...642A.210M}). We also observe a transition to weaker longitudinal fields as the spectral range narrows down around the line core.

Close to the limb, \citet{2019A&A...621A...1R} found longitudinal magnetic fields of $\sim$400~G in the photosphere decreasing to 200~G in the chromosphere of quiet-Sun areas. To estimate the latter, these authors applied the WFA to a spectral range of $\pm$1~\AA \, from the \ion{Ca}{II} 854.2~nm line center. At a heliocentric angle of 37$^{\circ}$, \citet{2020A&A...644A..43P} inferred $B_{\|}$ values of about 250~G outside a plage by performing STiC inversions of full-Stokes data in the $\lambda$8542 line (sampled within $\pm$880~m\AA \, from the line center). Also outside a plage but closer to the disk center, \citet{2020A&A...642A.210M} reported a weakening of $B_{\|}$ from $\sim -$600~G in the photosphere to $-$200~G in the low-mid chromosphere by applying a spatially constrained WFA method to the wings of the \ion{Na}{I} 589.6~nm line and the \ion{Ca}{II} 854.2~nm core ($\pm$110~m\AA). Recently, \citet{2021ApJ...911...41G} found maximum $B_{\|}$ values of 400--800~G in internetwork clusters in \ion{Fe}{I} 617.3~nm data that weaken by more than 200~G when applying the WFA to a spectral range of $\pm$200~m\AA \, in the \ion{Ca}{II} 854.2~nm line core.

We report $B_{\|}$ values of hG in quiet-Sun magnetic concentrations. Our mostly photospheric and chromospheric estimates in small and medium-sized structures are compatible with those previously found. However, our results for wider ranges across the line are stronger. Large concentrations also show stronger estimates, which is likely due to a departure of their physical properties from those typically found in the quiet Sun. Finally, we note that differences in the heliocentric angle may also lead to divergence of the results from those of other studies.

Furthermore, an interesting aspect of our study is the comparison of the photospheric and chromospheric estimates in concentrations of different size. Small concentrations usually host compact height variations of $B_{\|}$ exceeding 400–-500 G, where the maximum can be stronger than 800 G. Medium-sized and large structures show similar height variations but more extended over the concentrations. However, inspection of the average height variation of $B_{\|}$ reveals that small and medium-sized concentrations have stronger average height variation (of about 400 and 425 G, respectively) than large concentrations (of $\sim$300 G). Corroborating this result by analyzing other large concentrations will be of great interest because our sample only has two large structures, which were observed under poor seeing conditions. In addition, the height variations of $B_{\|}$ can be considered weaker or stronger than the chromospheric $B_{\|}$ estimated in each structure. In this regard, when compared to large concentrations, small and medium-sized concentrations display stronger height gradients in $B_{\|}$ than their chromospheric estimates. Consequently, large structures usually show better fits between the observed $V$ profiles and the scaled $I$ derivatives for wider spectral ranges.

Investigations of the stratification of the magnetic field with height in sunspots have revealed discrepancies depending on the approach used to determine the magnetic gradients (see the review of \citealt{2018SoPh..293..120B}). Undoubtedly, similar studies in the quiet Sun are also of great importance as they can shed light on the topology of its magnetic field. Nonetheless, the elusive detection of the weak polarization signals emerging from the quiet-Sun chromosphere hinders these investigations. This impasse will be overcome by using observations acquired with the next-generation solar telescopes, such as DKIST \citep{2020SoPh..295..172R} and EST \citep{2022A&A...666A..21Q}.

\section{Conclusions and summary}

Characterization of the magnetic field of the quiet-Sun chromosphere is essential for comprehending the chromospheric heating and the coupling of the magnetic field between the photosphere and the outer atmosphere. For this reason, we determined the longitudinal component of the magnetic field inferred in quiet-Sun magnetic concentrations depending on their size. Specifically, we applied the WFA to high-spatial resolution \ion{Ca}{II} 854.2~nm data acquired close to the disk center with SST/CRISP using different spectral ranges and temporally combined signals.

We find a correlation between the amplitude of $B_{\|}$ estimates and concentration size, with large structures hosting the largest values of $B_{\|}$. By restricting the spectral range to the line core, we infer chromospheric estimates that smoothly change from $\sim$50 at the edges to 150--500~G at the center of the concentrations. These values increase as the spectral range widens due to the influence of the photospheric contribution, which is especially strong compared to the chromospheric estimates in small and medium-sized concentrations. Consequently, the photospheric magnetic flux is three to four times greater than the chromospheric one in the large concentration and this difference can be as large as one order of magnitude in the small and medium-sized cases.

The effect of the height gradients in $B_{\|}$ on the reliability of the estimates given by the WFA depends on the concentration size and spectral range. Estimates for the inner-core range are usually reliable regardless of the concentration size, though they are affected by relative shifts between the $I$ and $V$ signals at some locations. In contrast, estimates for wider spectral ranges are generally unreliable in small and medium-sized structures.

Although differences in $B_{\|}$ and size are seen in small concentrations due to their dynamic behavior, signal addition usually leads to estimates akin to those at the original cadence for all spectral ranges. Therefore, we infer that the conditions along the considered formation regions may be stable with time in the studied structures and that signal addition has not warranted the detection of weaker signals. Moreover, signal addition reduces the noise of the chromospheric estimates and is not related to a better fulfillment of the WFA requisites when using wider ranges.

Finally, although deconvolving processes may enhance noise levels, data restored by destretching show similar results to those obtained when using the MOMFBD technique. However, results from the former are affected by the smearing caused by the partial correction of the data.

\begin{acknowledgements}
We thank the technical support provided by Dr.~\'{A}ngel de Vicente. We acknowledge the funding received from the European Research Council (ERC) under the European Union's Horizon 2020 research and innovation program (ERC Advanced Grant agreement No. 742265), as well as from the Spanish Ministerio de Ciencia, Innovaci\'on y Universidades through project PGC2018-102108-B-I00 and FEDER funds. This project has received funding also from the European Research Council (ERC) under the European Union's Horizon 2020 research and innovation programme (SUNMAG, grant agreement 759548). This paper is based on data acquired at the Swedish 1-m Solar Telescope, operated by the Institute for Solar Physics of Stockholm University in the Spanish Observatorio del Roque de los Muchachos of the Instituto de Astrofísica de Canarias. The Institute for Solar Physics is supported by a grant for research infrastructures of national importance from the Swedish Research Council (registration number 2017-00625). This research has made use of NASA's Astrophysical Data System.
\end{acknowledgements}

 \bibliographystyle{aa} 

\end{document}